\documentclass[12pt]{article}
\usepackage{amsmath,amssymb,enumerate,epsfig,caption2}

   {\end{list}}%

\makeatletter
\renewcommand{\section}{\setcounter{equation}{0}\@startsection
  {section}%
  {1}%
  {0pt}%
  {-1\baselineskip}%
  {0.4\baselineskip}%
  {\bfseries\large}}%
\renewcommand{\subsection}{\@startsection
  {subsection}%
  {2}%
  {0pt}%
  {-0.75\baselineskip}%
  {0.2\baselineskip}%
  {\bfseries}}%
\renewcommand{\subsubsection}{\@startsection
  {subsubsection}%
  {3}%
  {0pt}%
  {-0.5\baselineskip}%
  {0.1\baselineskip}%
  {\sc}}%
\makeatother

\makeatletter
 \newcommand\figcaption{\def\@captype{figure}\caption}
\makeatother

\def\Z{{\rm Z}}

\def\a{\alpha}
\def\b{\beta}
\def\d{\delta}

\def\e{\eta}
\def\la{\lambda}

\def\ka{\kappa}
\def\m{\mu}
\def\n{\nu}
\def\r{\rho}
\def\s{\sigma}

\def\th{\theta}

\def\id{{\rm{I}\!\rm{I}}}

\def\id3x{\int\!\! d^3\!\vec{x}}
\def\idx{\int\!\! d^4\!x}
\def\iDx{\int\!\! d^D\!x}

\def\rig>{\right>}

\newcommand{\bea}{\begin{eqnarray}}
\newcommand{\eea}{\end{eqnarray}}
\newcommand{\beann}{\begin{eqnarray*}}
\newcommand{\eeann}{\end{eqnarray*}}
\newcommand{\ba}{\begin{array}}
\newcommand{\ea}{\end{array}}

\def\g5{\gamma_{5}}

\def\idx3{\int\! d^{3}\!\vec{x}\,}
\def\idx{\int\! d^{4}\!x\,}

 \def\Dr {{\partial}_{\rho}}
 \def\Da {{\partial}_{\alpha}}
 \def\Db {{\partial}_{\beta}}
 \def\Dm {{\partial}_{\mu}}
 \def\Dn {{\partial}_{\nu}}

 \def\aa {a_{\alpha}}
 \def\ab {a_{\beta}}
 \def\am {a_{\mu}}

 \def\g {\gamma}

 \def\a {\alpha}
\def\b {\beta}
\def\r {\rho}
 \def\s {\sigma}

\begin{document}
\begin{titlepage}
\rightline{FTI/UCM 90-2006} \vglue 45pt

\begin{center}

{\Large \bf The noncommutative $U(1)$ Higgs-Kibble model in the enveloping-algebra formalism and its renormalizability.}\\
\vskip 1.2 true cm {\rm C.P. Mart\'{\i}n}\footnote{E-mail:
carmelo@elbereth.fis.ucm.es},
D. S\'anchez-Ruiz\footnote{E-mail: domingo@toboso.fis.ucm.es}
and  C. Tamarit\footnote{E-mail: ctamarit@fis.ucm.es}
\vskip 0.3 true cm {\it Departamento de F\'{\i}sica Te\'orica I,
Facultad de Ciencias F\'{\i}sicas\\
Universidad Complutense de Madrid,
 28040 Madrid, Spain}\\
\vskip 0.75 true cm

\vskip 0.25 true cm

{\leftskip=50pt \rightskip=50pt \noindent We discuss the
renormalizability of the  noncommutative $U(1)$ Higgs-Kibble model
formulated within the enveloping-algebra approach. We consider
both the phase of the model with unbroken gauge symmetry and  the
phase with spontaneously broken gauge symmetry. We show that against all
odds the gauge sector of the model is always one-loop
renormalizable at first order in $\theta^{\m\n}$, perhaps, hinting
at the existence of a new symmetry of the gauge sector of the
model. However,  we also show that the matter sector of the model
is non-renormalizable whatever the phase.

\par }
\end{center}

\vspace{20pt} \noindent
 \vfill
\end{titlepage}


\setcounter{page}{2}

\section{Introduction}

At present, there is only one available framework to formulate
gauge theories in noncommutative space-time for an arbitrary
simple gauge group in an arbitrary representation. This very
framework is the only known formalism where one may have fields
with arbitrary $U(1)$ charge. The formalism we are referring to
was introduced in refs.~\cite{Madore:2000en,Jurco:2000ja}
and~\cite{Jurco:2001rq} and led  to the formulation of the
noncommutative standard model~\cite{Calmet:2001na} and some Grand
Unification models~\cite{Aschieri:2002mc}. Some phenomenological
implications of these models have been studied
recently~\cite{Melic:2005hb, Melic:2005su, Haghighat:2005jy,
MohammadiNajafabadi:2006iu, Alboteanu:2006hh, Buric:2006nr}, but quite a lot of
work remains to be done in view of the coming of the LHC.

As is well known, the Seiberg-Witten map plays a central role in
the framework of refs.~\cite{Madore:2000en,Jurco:2000ja}
and~\cite{Jurco:2001rq}. Indeed, the noncommutative gauge fields
are defined in terms of the ordinary fields by means of the formal
series expansion in powers of the noncommutative matrix parameter
$\theta^{\m\n}$ that implements the Seiberg-Witten map. The
noncommutative gauge fields do not thus belong, in general, to the
Lie algebra of the gauge group but are valued in the enveloping
algebra --this is why the formalism is called the
enveloping-algebra formalism-- of that Lie algebra. This is quite
at variance with the alternative approach to model building in
noncommutative space-time employed in
refs.~\cite{Chaichian:2001py, Khoze:2004zc, Abel:2005rh}
and~\cite{Arai:2006ya}.

The renormalizability of some noncommutative field theory models
constructed within the enveloping-algebra formalism has been
studied in a number of papers: see refs.~\cite{Bichl:2001cq,
Wulkenhaar:2001sq, Buric:2002gm, Buric:2004ms, Buric:2005xe,
Buric:2006wm}. In all these papers, and throughout this one, it is
assumed that both the quantization procedure and the
renormalization program deal  with the  1PI functions of the
ordinary fields that define the noncommutative fields via the
Seiberg-Witten map. The reader is referred to
ref.~\cite{Calmet:2006zy} for an alternative interesting proposal.
The models whose UV divergences have been worked out in
refs.~\cite{Bichl:2001cq, Wulkenhaar:2001sq, Buric:2002gm,
Buric:2004ms, Buric:2005xe, Buric:2006wm} only have $U(1)$ and/or
$SU(N)$ gauge fields and Dirac fermions in the fundamental
representation. It turns out that at first order in
$\theta^{\m\n}$, and against all odds, the one-loop UV divergences
of the Green functions that only involve gauge fields in the
external legs are renormalizable in the models that have and have
not Dirac fermions. This is quite a surprising result since, as
already pointed out in ref.~\cite{Wulkenhaar:2001sq}, BRST
invariance on its own cannot account for it, thus hinting at the
existence of an as yet unveiled symmetry of the noncommutative
gauge sector of these models. The result in question is even more
surprising if one takes into account that the Green functions that
carry fermion fields in the external legs cannot all be
renormalized, thus rendering nonrenormalizable in the
enveloping-algebra approach all the noncommutative models studied
so far.

The main purpose of this paper is to see whether the results
summarized in the previous paragraph also hold when the matter
fields are not Dirac fermions but scalar fields --let us recall
that the Higgs field is a key ingredient of the Standard Model.
The simplest model that captures some of the features of the
noncommutative Standard Model and includes both gauge fields and
scalar fields is the noncommutative $U(1)$ Higgs-Kibble model.
This model has a phase where the $U(1)$ symmetry is spontaneously
broken and has also a phase where the $U(1)$ symmetry is not
broken. The renormalization properties of the noncommutative
$U(1)$ Higgs-Kibble model have never been studied when formulated
within the enveloping-algebra formalism, although they have been
analyzed within the standard noncommutative field theory formalism
--see ref.~\cite{Petriello:2001mp} for the $U(1)$ Higgs-Kibble
model and refs.~\cite{Campbell:2000ug, Liao:2001uv, Liao:2002tv, 
RuizRuiz:2002hh} for other models with spontaneous symmetry breaking.

The computation we are about to sketch is quite a daunting
one since it demands the calculation of 94  1PI Feynman
diagrams to tell whether the model is renormalizable in the phase
with no symmetry breaking. In this phase, we discuss both the
massive and massless cases. To deal with such a large number of
Feynman diagrams we have used the algebraic manipulation package
$\tt Mathematica$~\cite{mathematica}. Then, we shall  use the
results obtained in the phase with no symmetry breaking to analyze
the renormalizablity of the model in the phase with spontaneous
symmetry breaking.

The layout of this paper is as follows. In section 2, we define
the classical noncommutative $U(1)$ Higgs-Kibble model and work
out the action up to first order in $\theta^{\m\n}$. The
renormalizability of the model in the phase with unbroken gauge
symmetry is discussed in section 3. In section 4, we analyze the
renormalizability of the model in the phase with spontaneous
symmetry  breaking. A summary of the results obtained in the paper
is given in section 5. The Feynman rules and Feynman diagrams
quoted in  the paper can be found in the appendix.

\section{The action. The Seiberg-Witten map}

As it was stated in the introduction, our noncommutative field theory model will be the $U(1)$ Higgs-Kibble model.
The model contains a noncommutative $U(1)$ gauge field $A_\mu$ and  a
noncommutative complex scalar field $\Phi$ coupled to $A_\mu$.
The classical action of the model in terms of the noncommutative fields reads
\begin{equation}
 S_{\rm class}=\idx-\frac{1}{4} F_{\m\n}\star F^{\m\n}+(D_\mu\Phi)^* \star D^\mu \Phi-\mu^2\Phi^*\star\Phi-\frac{\lambda}{4}(\Phi^*\star\Phi)^2,
\label{Snc}
\end{equation}
where
\begin{equation*}
 D_\m\Phi=\Dm\Phi-i e A_\m \star\Phi,\quad F_{\m\n}=\Dm A_\n-\Dn A_\m-i e[A_\m,A_\n]_\star.
\end{equation*}
$e$ denotes the coupling constant of the gauge interaction. $\mu$
is the mass parameter and $\lambda$ stands for the coupling
constant of the scalar self-interaction, which we shall take to be
positive.

In the enveloping-algebra approach the noncommutative fields are
defined in terms of the ordinary fields --the ordinary $U(1)$
gauge field $a_\mu$ and the ordinary complex scalar $\phi$ with
$U(1)$ charge $e$-- by means of the Seiberg-Witten map.  It is the
ordinary fields $a_\mu$ and $\phi$ that will be chosen as the field
variables to be used to first quantize and then renormalize the theory.

At first order in $h\theta^{\mu\nu}$, the most general Seiberg-Witten map reads
\begin{equation}
 \begin{array}{l}
A_\mu=\am-\frac{e h}{2}\theta^{\alpha\beta}a_\alpha (2\partial_\beta\am-\Dm\ab)+h \partial_\mu H+hS_\mu+O(h^2),\\
\Phi=\phi-\frac{e h}{2}\theta^{\alpha\beta}a_\alpha\partial_\beta\phi+i h H \phi+h F+O(h^2),\\
H=x_1\, \theta^{\alpha\beta}\Da\ab\\
S_\mu=\kappa_1\,\theta^{\alpha\beta}\Dm
f_{\alpha\beta}+\kappa_2\,{\theta_\mu}^\beta \partial^\nu f_{\n\beta}+e \ka_3\,{\theta_\m}^\n\Dn(\phi^*\phi)+i e\ka_4{\theta_\m}^\n(D_\n\phi^*\phi-\phi^*D_\n\phi) ,\\
F=e\kappa_5\theta^{\a\b}f_{\a\b}\phi,
\end{array}
\label{SWmap}
\end{equation}
which has five parameters --four real parameters
$\kappa_1,\,\kappa_2,\,\kappa_3,\, \kappa_4$ and a complex
parameter $\kappa_5$-- labelling the ambiguity associated with
field redefinitions. The  real parameter $x_1$ parametrizes a
gauge transformation of the fields.

For convenience, we introduce next the following basis
$\{t_i\}_{i=1,\dots, 9}$ of independent, modulo total derivatives,
and gauge invariant monomials that are of order one in $h\theta^{\m\n}$
and have mass dimension equal to four:
\begin{equation}
\begin{array}{lll}
 t_1=\theta^{\a\b}f_{\a\b}f_{\r\s}f^{\r\s}     & t_2=\theta^{\a\b}f_{\a\r}f_{\b\s}f^{\r\s} & t_3=\theta^{\a\b}\phi^*\phi\,\square f_{\a\b}\\
 t_4= \theta^{\a\b}(D_\r\phi)^*\phi\,\partial^\r f_{\a\b}   & t_5=\theta^{\a\b}(D_\a\phi)^*\phi\,\partial_\r {f_\b}^\r  & t_6=\theta^{\a\b}(D_\r\phi)^*D^\r\phi f_{\a\b} \\
 t_7=\theta^{\a\b}(D_\a\phi)^*D^\r\phi f_{\b\r}      & t_8=\theta^{\a\b}f_{\a\b}\,(\phi^*\phi)^2  & t_9=\mu^2 \theta^{\a\b}f_{\a\b}\,\phi^*\phi.\\
\end{array}
\label{tbasis}
\end{equation}
    Substituting first the Seiberg-Witten map of eq.~\eqref{SWmap} in the action in eq.~\eqref{Snc} and then expanding in powers of $h\theta^{\m\n}$, one obtains
\begin{equation}
 S_{\rm class}=S^{(0)}+hS^{(1)}+O(h^2),
\label{Sexp}
\end{equation}
where $S^{(0)}$ is the ordinary classical contribution,
\begin{equation}
 S^{(0)}=\idx-\frac{1}{4} f_{\m\n}f^{\m\n}+(D_\mu\phi)^*  D^\mu \phi-\mu^2\phi^*\phi-\frac{\lambda}{4}(\phi^*\phi)^2,
\label{Sord}
\end{equation}
--now, $D_\mu=\partial_\m-iea_\mu$-- and $S^{(1)}$ has the
following form in terms of the $t_i$s defined in
eq.~\eqref{tbasis}:
\begin{equation}
S^{(1)} =\idx \frac{e}{8}\,t_1-\frac{e}{2}\,t_2+e C_3\,t_3+e C_4\,t_4+e C_5\,t_5+eC_6t_6+eC_7t_7+eC_8t_8+e\mu^2C_9t_9,\
\label{S1}
\end{equation} where
\begin{equation}
\begin{array}{lll}
 C_3=-\ka_5^*-i \kappa_1-\frac{i}{2}\,(\kappa_2+\ka_4)+\frac{1}{2}\,\kappa_3 & C_4=-\frac{1}{4}+2i\,\text{Im}\ka_5-2i\kappa_1 & C_5=-\frac{1}{2}+2i(\kappa_2+\ka_4)\\
  C_6=-\frac{1}{4}+2\,\text{Re}\ka_5 & C_7=-1 & C_9=\frac{1}{4}-2\,\text{Re}\ka_5\\
C_8=\frac{e^2}{2}\,\kappa_3-\lambda(\text{Re}\ka_5-\frac{1}{16}).
 \end{array}
\label{coeffS1}
\end{equation}



\section{The model in the phase with unbroken symmetry}

In this section we shall show that the gauge sector of the model with unbroken
gauge symmetry is one-loop renormalizable at first order in $h\theta^{\m\n}$,
and that the matter sector is not renormalizable.

\subsection{Feynman rules and one-loop UV divergences}

In the case at hand $\mu^2\geq0$, so that the classical vacuum of the theory is the trivial field configuration $\phi=0$ and $a_\mu=0$. To quantize the theory at first order
in $h\theta^{\m\n}$, we shall add to the classical action in eq.~\eqref{Sexp}
the gauge-fixing, $S_{\rm gf}$, and ghost, $S_{\rm gh}$, terms, to obtain
\begin{equation}
S=S_{\rm class}+S_{\rm gf}+S_{\rm gh}=S^{(0)}+hS^{(1)}+S_{\rm gf}+S_{\rm gh},
\label{quact}
\end{equation}
where
\begin{equation*}
S_{\rm gf}=\idx-\frac{1}{2\xi}\,(\Dm a^\m)^2,\quad S_{\rm gh}=\idx \bar{c}\partial^2\,c.
\end{equation*}
Recall that it is the ordinary fields $a_\m$ and $\phi$ that furnish the
field variables to be used to carry out the quantization process: in the path integral we
shall integrate over $a_\m$ and $\phi$. Notice that for our choice of gauge
fixing, the ghost fields, $c$ and $\bar{c}$, do not couple either to $a_\m$
or to $\phi$, and hence we will  dispose of them.

The Feynman rules that the action in eq.~\eqref{quact} gives rise
to are depicted in figure 1 of the appendix, where the following
notation is used for propagators and vertices:

{\bf Propagators}
\begin{equation*}
\begin{array}{l}
 a_\mu\longrightarrow G^{\mu\n}(k)=\frac{i}{k^2+i\varepsilon}\Big[-g^{\m\n}+(1-\xi)\frac{k^\mu k^\nu}{k^2}\Big]\\
\phi \longrightarrow G(k)=\frac{i}{k^2-m^2+i\varepsilon}
\end{array}
\end{equation*}\par
{\bf Ordinary vertices}
\begin{equation*}
\begin{array}{lll}
 \Gamma^{t(0)\,\m}_{(1,1)}[r;p,q]=ie(p^\m+q^\m) &  \Gamma^{t(0)\,\m\n}_{(2,1)}[r,s;p,q]=2 i e^2 g^{\m\n} & \Gamma^{t(0)}_{(0,2)}[r,s;p,q]=-i\lambda
\end{array}
\end{equation*}
\par
{\bf Noncommutative vertices}

\begin{equation}
\begin{array}{l}
 \Gamma^{t(1)\,\m}_{(1,1)}[r;p,q]=\\
e\Big[-2 C_3 \th^{\a\m}(p-q)_\a(p-q)^2+2C_4\th^{\a\m}(p-q)_\a(p-q)\cdot p-C_5\th^{\a\b}p_\a q_\b(p-q)^\m\\
\phantom{e\Big[}-C_5\th^{\a\m}p_\a(p-q)^2+2C_6\th^{\a\m}(p-q)_\a p\cdot q+C_7\th^{\a\b}p_\a(p-q)_\b q^\m-C_7\th^{\a\m}p_\a q\cdot(p-q)\\
\phantom{e\Big[}+2C_9\m^2\th^{\a\m}(p-q)_\a\Big]\\
\Gamma^{t(1)\,\m\n\e}_{(3,0)}[p,q,r]=\\
-\frac{e}{2}\Big[2\th^{\a\m}p_\a(q\cdot r g^{\n\e}-q^\e r^\n)+2\th^{\a\n}q_\a(r\cdot p g^{\e\m}-r^\m p^\e)+2\th^{\a\e}r_\a(p\cdot q g^{\m\n}-p^\n q^\m)\Big]\\
+e\Big[\th^{\a\b}p_\a q_\b(r^\m g^{\n\e}-r^\n g^{\e\m})+\th^{\a\b}p_\a r_\b(q^\m g^{\n\e}-q^\e g^{\m\n})+\th^{\a\b}q_\a r_\b(p^\n g^{\m\e}-p^\e g^{\n\m})\\
\phantom{e\Big[}-\th^{\a\n}p_\a q^\e r^\m-\th^{\a\e}p_\a r^\n q^\m-\th^{\a\e}q_\a r^\m p^\n-\th^{\a\m}q_\a p^\e r^\n -\th^{\a\m}r_\a p^\n q^\e-\th^{\a\n}r_\a q^\m p^\e\\
\phantom{e\Big[}-q_\b p\cdot r(\th^{\m\b}g^{\n\e}+\th^{\e\b}g^{\n\m})-r_\b p\cdot q(\th^{\m\b}g^{\e\n}+\th^{\n\b}g^{\e\m})-p_\b q\cdot r(\th^{\n\b}g^{\m\e}+\th^{\e\b }g^{\m\n})\\
\phantom{e\Big[}+\th^{\m\n}(p\cdot r q^\e-q\cdot r p^\e)+\th^{\m\e}(p\cdot q r^\n-r\cdot q p^\n)+\th^{\n\e}(q\cdot p r^\m-r\cdot p q^\m)
\Big]\\
\Gamma^{t(1)\,\m\n}_{(2,1)}[r,s;p,q]=\\
e^2\Big[C_4(2\th^{\a\n}s^\m s_\a+2\th^{\a\m}r^\n r_\a)+C_5(\th^{\m\b}s^\n s_\b+\th^{\n\b}r^\m r_\b-\th^{\m\n}(s^2-r^2))\\
\phantom{e\Big[}+C_6(2\th^{\a\n}(q+p)^\m s_\a+2\th^{\a\m}(q+p)^\n r_\a)+C_7(\th^{\m\b}s_\b q^\n-\th^{\m\n} q\cdot s+\th^{\a\b}p_\a s_\b g^{\m\n}-\th^{\a\n}p_\a s^\m\\
\phantom{e\Big[}+\th^{\n\b}r_\b q^\m+\th^{\m\n} q\cdot r+\th^{\a\b}p_\a r_\b g^{\m\n}-\th^{\a\m}p_\a r^\n)\Big]\\
  \Gamma^{t(1)\,\m\n\r}_{(3,1)}[r,s,t;p,q]=\\
e^3\Big[4C_6(\th^{\a\r}g^{\m\n}t_\a+\th^{\a\n}g^{\m\r}s_\a+\th^{\a\m}g^{\n\r}r_\a)+C_7(\th^{\m\b}g^{\n\r}(t+s)_\b+\th^{\n\b}g^{\r\m}(r+t)_\b\\
\phantom{e^3\Big[}+\th^{\r\b}g^{\m\n}(s+r)_\b-\th^{\m\r}t^\n-\th^{\m\n}s^\r-\th^{\n\m}r^\r-\th^{\n\r}t^\m-\th^{\r\n}s^\m-\th^{\r\m}r^\n)\Big]\\
 \Gamma^{t(1)\,\m}_{(1,2)}[t;p,q;r,s]=8e C_8\th^{\a\m}t_\a.
 \end{array}
\label{feynmannossb}
\end{equation}
$C_i$, $i=1\dots 9$, have been  given in eq.~\eqref{S1}.

Now, using the fact that the BRST transformations of $a_\mu$ and
$\phi$ read $sa_\m=\partial_\m c$ and $s\phi=ie\,\phi\, c$,
respectively, it is not difficult to conclude that in dimensional
regularization the pole part of the one-loop 1PI functional,
$\Gamma[a_\mu,\phi]_{\rm pole}^{\rm
 one-loop}$, must be gauge invariant. Hence, up
to first order in $h\theta^{\mu\n}$ this functional should read
\begin{equation}
 \Gamma[a_\mu,\phi,\phi^*]_{\rm pole}^{\rm
 one-loop}=\Gamma^{(0)}[a_\mu,\phi,\phi^*]_{\rm pole}^{\rm
 one-loop}\,+\,h\,
\Gamma^{(1)}[a_\mu,\phi,\phi^*]_{\rm pole}^{\rm one-loop},
\label{1PIfun}
\end{equation}
where
\begin{equation}
\begin{array}{l}
\Gamma^{(0)}[a_\mu,\phi,\phi^*]_{\rm pole}^{\rm
 one-loop}= \idx\;-\frac{w_1}{4} f_{\m\n} f^{\m\n}+w_2\,(D_\mu\phi)^*
\star D^\mu \phi-
w_3\,\mu^2\phi^* \phi-w_4\,\frac{\lambda}{4}(\phi^* \phi)^2,\\
\Gamma^{(1)}[a_\mu,\phi,\phi^*]_{\rm pole}^{\rm
 one-loop}=\idx\;\frac{e}{8}\,z_1\,t_1-\frac{e}{2}\,z_2\,t_2+\sum_{i=3}^8 e z_i\,t_i+e\mu^2z_9t_9.
\end{array}
\label{wscoeff}
\end{equation}
The $t_i$s, $i=1\dots 9$, are the nine monomials in
eq.~\eqref{tbasis}, and $w_i$, $i=1\dots 4$,  and  $z_i$,
$i=1\dots 9$, stand for  coefficients that are simple poles in
$\epsilon=D/2-2$.

Let $\Gamma_{(m,n)}^{\mu_1\dots\mu_m}[x_k;y_l;z_l]$ denote the 1PI
Green function corresponding to $m$ $a_\mu(x)$ fields, $n$
$\phi^*(y)$ and $n$ $\phi(z)$ fields. Ignoring the tree-level
ghost contribution, we have that the 1PI functional reads
\begin{equation}
 \Gamma[a_\mu,\phi,\phi^*]=\sum_{m,n}\frac{1}{m!(n!)^2}\int\prod_{k=1}^m\prod_{l=1}^n d^Dx_k d^Dy_l d^Dz_l
 \Gamma_{(m,n)}^{\mu_1\dots\mu_m}[x_k;y_l;z_l] a_{\mu_k}(x_k)\phi^*(y_l)\phi(z_l).
\label{fornotation}
\end{equation}

The computation of the $w_i$s, $i=1,\dots,4$, in eq.~\eqref{wscoeff} is
a standard exercise in introductory courses to renormalization
theory, so  we will just quote the result:
\begin{equation}
w_1=-\frac{e^2}{48\pi^2\epsilon},\;w_2=\frac{e^2}{16\pi^2\epsilon}\,(3-\xi),\,w_3=-\frac{e^2\xi-\lambda}{16\pi^2\epsilon},\,w_4=\frac{1}{32\pi^2\epsilon}\,\Big[24
\frac{e^4}{\lambda}-4 e^2\xi+5\lambda\Big]
\label{wpole}.
\end{equation}
The computation of the $z_i$s, $i=1,\dots, 9$, in eq.~\eqref{wscoeff}
is, though, a very lengthy and involved computation since the pole
part of a large number of topologically inequivalent diagrams --94
altogether-- with a single noncommutative vertex --which is in
general a long expression-- must  be worked out. It turns out that
to obtain all the $z_i$s  one must evaluate the pole part of the
one-loop contributions to
$\Gamma^{\m\n\r}_{(3,0)},\,\Gamma^{\m}_{(1,1)}$ and
$\Gamma^\m_{(1,2)}$ that are linear in $\theta^{\m\n}$ --see
eq.~\eqref{fornotation} for notation. Let us next display the
values of these one-loop pole parts that we shall denote,
respectively, by
$\Gamma^{(1)\m\n\r}_{(3,0)}[p_1,p_2,-p_1-p_2]_{\rm pole}^{\rm
one-loop}$, $\Gamma^{(1)\m}_{(1,1)}[p_1-p_2;p_1,p_2]_{\rm
pole}^{\rm one-loop}$ and
$\Gamma^{(1)\m}_{(1,2)}[p_1+p_3-p_2-p_4;p_1,p_2,p_3,p_4]_{\rm
pole}^{\rm one-loop}$:

 {\bf The $aaa$ 1PI Green function $\Gamma^{\m\n\r}_{(3,0)}$}.

There are 4 topologically inequivalent diagrams --see figure 2 in
the appendix-- contributing to the pole part of this Green
function at first order in $h\theta^{\m\n}$, and they  lead to the
following result:
\begin{equation}
\Gamma^{(1)\m\n\r}_{(3,0)}[p_1,p_2,-p_1-p_2]_{\rm pole}^{\rm
one-loop}=-\frac{e^2}{48\pi^2\epsilon}\,
\Gamma^{t(1)\m\n\r}_{(3,0)}[p_1,p_2,-p_1-p_2],
\label{aaauv}
\end{equation}
where $\Gamma^{t(1)\m\n\r}_{(3,0)}$ is the tree-level contribution
given in eq.~\eqref{feynmannossb} coming from the contributions
$t_1$ and $t_2$ to $S_{\rm class}$ --see eqs.~\eqref{Sexp}
and~\eqref{S1}.

 {\bf The $a\phi^*\phi$ 1PI Green function $\Gamma^{\m}_{(1,1)}$}.

    The pole parts of the 11 topologically inequivalent diagrams in
    figure 3 of the appendix are to be computed, to obtain the following
    answer:
\begin{equation}
\Gamma^{(1)\m}_{(1,1)}[p_1-p_2;p_1,p_2]_{\rm pole}^{\rm one-loop}=
\Gamma^{t(1)\,\m}_{(1,1)}[\Delta_3,\Delta_4,\Delta_5,\Delta_6,\Delta_7,\Delta_9;\,
p_1-p_2;p_1,p_2], \label{1a2phi}
\end{equation}
where
$\Gamma^{t(1)\,\m}_{(1,1)}[\Delta_3,\Delta_4,\Delta_5,\Delta_6,\Delta_7,\Delta_9;\,
r;p,q]$ is obtained from $\Gamma^{t(1)\,\m}_{(1,1)}[r;p,q]$ in
eq.~\eqref{feynmannossb} by replacing $C_i$ with $\Delta_i$,
$i=3,4,5,6,7$ and $9$, where
\begin{equation}
 \begin{array}{l}
 \Delta_3=\frac{e^2}{192\pi^2\epsilon}[11+26 C_4-13 C_5-26 C_6+13 C_7-4C_3(1+3\xi)-\lambda(-12C_3+6C_4-3C_5-6C_6\\
\phantom{\Delta_3=\frac{e^2}{192\pi^2\epsilon}[}+2C_7)]\\
  \Delta_4=\frac{e^2}{32\pi^2\epsilon}[2-C_5+2C_7+2C_4(4-\xi)]\\
\Delta_5=\frac{e^2}{16\pi^2\epsilon}[2(1+C_7)+C_5(3-\xi)]\\
\Delta_6=\frac{e^2}{32\pi^2\epsilon}[-1+C_7+2C_6(4-\xi)]\\
\Delta_7=\frac{e^2}{16\pi^2\epsilon}[4+C_7(7-\xi)]\\
\Delta_9=\frac{1}{32\pi^2\epsilon}[8C_8+e^2(-8C_3+4C_4-2C_5-2C_6+2C_7+2C_9(2-\xi))-\lambda(-4C_6+C_7-2C_9)].
\end{array}
\label{deltais}
\end{equation}
The constants $C_i$, $i=3,4,5,6,7$ and $9$, are defined in
eq.~\eqref{coeffS1}.

 {\bf The $a\phi^*\phi\phi^*\phi$ 1PI Green function $\Gamma^{\m}_{(1,2)}$}.

    Let $\Gamma^{t(1)\m}_{(1,2)}[\Delta_8;\,t;p,q;r,s]$ denote
    $\Gamma^{t(1)\m}_{(1,2)}[t;p,q;r,s]$ in eq.~\eqref{feynmannossb}, once $C_8$ is replaced
    with $\Delta_8$. Then, the computation of the pole part of the 79 topologically inequivalent diagrams in figure 4 of the appendix leads
    to the following equality:
\begin{equation}
\Gamma^{(1)\m}_{(1,2)}[p_1+p_3-p_2-p_4;p_1,p_2;p_3,p_4]_{\rm
pole}^{\rm one-pole}=
\Gamma^{t(1)\m}_{(1,2)}[\Delta_8;\,p_1+p_3-p_2-p_4;p_1,p_2;p_3,p_4], \label{1a4phi}
\end{equation}
where
\begin{equation}
\begin{array}{l}
\Delta_8=\frac{1}{256\pi^2\epsilon}\,[-4e^4(-12+4C_3 -2 C_4+C_5+26C_6-8C_7)+5\lambda^2(4C_6-C_7)\\
\phantom{\Delta_8=\frac{1}{256\pi^2\epsilon}\,[}-8e^2\lambda(8C_3-4C_4+2C_5+C_6-C_7)+32e^2C_8(2-\xi)+80\lambda
C_8].
\end{array}
\label{delta8}
\end{equation}

Taking into account eqs.~\eqref{aaauv},~\eqref{1a2phi} and~\eqref{1a4phi}, one
concludes that the $z_i$s, $i=1,\dots,9$, in eq.~\eqref{wscoeff} are given by the
following equalities:
\begin{equation}
z_1=z_2=-\frac{e^2}{48\pi^2\epsilon},\,z_i=\Delta_i,\;\forall\,i=3\dots 9,
\label{zisvalues}
\end{equation}
where the $\Delta_i$s, $i=3\dots 9$, are given in eqs.~\eqref{deltais} and~\eqref{delta8}.

\subsection{One-loop renormalization}

Let us assume that the fields and parameters of the action in eq.~\eqref{quact} are the bare fields and parameters of the model. Then, as usual, we shall say
that the model is one-loop multiplicatively renormalizable at first order in $h\theta^{\m\n}$, if the free coefficients of the counterterm
action obtained by introducing the following renormalizations of the
fields and parameters of the action in eq.~\eqref{quact}
\begin{equation}
 \begin{array}{lll}
 a_\m=Z_a^{1/2}a^R_\m   &   \phi=Z_\phi^{1/2}\phi^R & e=Z_e e^R \\
  \m=Z_\m^{1/2}\m^R & \lambda=Z_\lambda \lambda^R & \xi=Z_\xi\xi^R\\
  \theta=Z_\theta \theta^R  & \kappa_i=\kappa_i^R+\delta\kappa_i,\,i=1,2,3,4 & \text{Re}\kappa_5=\text{Re}\kappa_5^R+\delta\text{Re}\kappa_5^R\\
\text{Im}\kappa_5=\text{Im}\kappa_5^R+\delta\text{Im}\kappa_5^R
\end{array}
\label{mulren}
\end{equation}
can be chosen to cancel the UV divergences of the 1PI functional given in
eqs.~\eqref{1PIfun},~\eqref{wscoeff},\\~\eqref{wpole} and~\eqref{zisvalues}.

Let $\delta Z_a=Z_a-1$, $\delta Z_\phi=Z_\phi-1$, $\delta Z_e=Z_e-1$,
$\delta Z_\mu=Z_\mu-1$, $\delta Z_\lambda=Z_\lambda-1$, $\delta Z_\xi=Z_\xi-1$
and $\delta Z_\theta=Z_\theta-1$. Then,
the multiplicative renormalization in eq.~\eqref{mulren}, when applied to
the action in eq.~\eqref{quact}, yields the following one-loop counterterm
action up to first order in $h\theta^{\mu\n}$:
\begin{equation*}
S_{\rm ct}\,=\,S^{(0)}_{\rm ct}\,+\,h\,S^{(1)}_{\rm ct},
\end{equation*}
where
\begin{equation}
\begin{array}{l}
 S^{(0)}_{\rm ct}=\idx-\frac{1}{4}\delta Z_a f_{\m\n}f^{\m\n}+
\delta Z_\phi (\Dm\phi)^*  \partial^\mu \phi-(\delta Z_\phi+ \delta Z_\mu)\mu^2\phi^*\phi\\
\phantom{S^{(0)}_{\rm ct}=}-i e(\delta Z_\phi+\delta Z_e+\frac{1}{2} \delta Z_a )\Dm \phi^*a^\mu\phi
+i e(\delta Z_\phi+\delta  Z_e+\frac{1}{2}\delta Z_a )\phi^*a^\mu\Dm\phi\\
\phantom{S^{(0)}_{\rm ct}=}+e^2(2\,\delta Z_e + \delta Z_a+\delta Z_\phi)\phi^*\phi\am a^\mu
-\frac{\lambda}{4}(\delta Z_\lambda+2 \delta Z_\phi)(\phi^*\phi)^2
-\frac{1}{2\xi}\,(\delta Z_a- \delta Z_\xi)(\Dm a^\mu)^2,\\[10pt]
 S^{(1)}_{\rm ct}=\idx \frac{e}{8}\, (\delta Z_\theta +\delta Z_a) t_1-\frac{e}{2} (\delta Z_\theta+\delta Z_a)  t_2+e\sum_{i=3}^7[\delta C_i+C_i(\delta Z_\theta +\delta Z_\phi)] t_i\\
\phantom{S^{(1)}=}+e[\delta C_8+C_8(\delta Z_\theta +2\delta Z_\phi)] t_8+e \mu^2[\delta C_9+C_9(\delta Z_\theta+\delta Z_\mu+\delta Z_\phi)]t_9.
\end{array}
\label{order1countterms}
\end{equation}
 In the previous equation the fields $a_\mu$ and $\phi$ and the parameters $\mu^2, e,\lambda$ and $\kappa_i$, $i=1,\dots 4$ are, respectively, the renormalized
fields and parameters of eq.~\eqref{mulren}. We have suppressed
the superscript ``$R$'' to make the notation simpler. To simplify the expression for $ S^{(1)}_{\rm ct}$, the identity $\delta Z_a=-2 \delta Z_e$, which is
a consequence of the BRST invariance of the theory, has been used. Notice that
as a consequence of the identities in eq.~\eqref{coeffS1} the $\delta
C_i$s in eq.~\eqref{order1countterms} are defined by following
equalities:
\begin{equation}
 \begin{array}{ll}
 \delta C_3=-\delta\text{Re}\ka_5+i\delta\text{Im}\ka_5-\frac{i}{2}\,(\delta\kappa_2+\delta\ka_4)+\frac{1}{2}\,\delta\kappa_3-i\delta\kappa_1 &
\delta C_4=2i\delta\text{Im}\ka_5-2i\delta\kappa_1\\
\delta C_5=2i(\delta\kappa_2+\delta\ka_4) & \delta C_6=2\delta\text{Re}\ka_5\\
\delta C_7=0 & \delta C_9=-2\delta\text{Re}\ka_5\\
\delta C_8=\frac{1}{2}(2 e^2 \kappa_3 \delta Z_e+e^2\delta\kappa_3)-\lambda \delta Z_\lambda(\text{Re}\ka_5-\frac{1}{16})-\lambda\delta\text{Re}\ka_5.
\end{array}
\label{deltaCs}
\end{equation}

Of course, $\delta Z_a$, $\delta Z_\phi$, $\delta Z_e$,
$\delta Z_\mu$, $\delta Z_\lambda $ and  $\delta Z_\xi$ are the same as in
the ordinary model, and in the MS scheme they read
 \begin{equation}
\begin{array}{l}
 \delta Z_a=\frac{e^2}{48\pi^2\epsilon}=-2\delta Ze=\delta Z_\xi,\quad
\delta Z_\phi=-\frac{e^2(3-\xi)}{16 \pi^2\epsilon},\\[8pt]
\delta Z_\mu=\frac{3 e^2-\lambda}{16\pi^2\epsilon}\quad
\delta Z_\lambda=-\frac{1}{32\pi^2\epsilon}\,\Big[24 \frac{e^4}{\lambda}-12 e^2+5\lambda\Big].\\
\end{array}
\label{ordZ}
\end{equation}
 Next, in the MS scheme, $\delta Z_{\theta}$ and $\delta C_i$, $i=3,\dots,9$, of $S^{(1)}_{\rm ct}$ in eq.~\eqref{order1countterms}
must be chosen --were it possible-- so that the sum
$\Gamma^{(1)}[a_\mu,\phi,\phi^*]_{\rm pole}^{\rm one-loop}+S^{(1)}_{\rm ct}$ vanishes. $\Gamma^{(1)}[a_\mu,\phi,\phi^*]_{\rm pole}^{\rm one-loop}$ is given
in eq.~\eqref{wscoeff} and the values of its coefficients --the $z_i$s--  are summarized in eq.~\eqref{zisvalues}. We thus conclude that $\delta Z_{\theta}$
must satisfy the following equalities:
\begin{equation}
-z_1=\delta Z_\theta +\delta Z_a,\quad -z_2=\delta Z_\theta +\delta Z_a,
\label{eqztheta}
\end{equation}
whereas for $\delta C_i$, $i=3,\dots,9$ the following set of equations must
hold:
\begin{equation}
\begin{array}{l}
-z_i\,=\,\delta C_i\,+\, C_i(\delta Z_\theta +\delta Z_\phi),\quad i=3,4,5,6,7\\
-z_8=\delta C_8\,+\, C_8(\delta Z_\theta +2\,\delta Z_\phi),\\
-z_9=\delta C_9\,+\, C_9(\delta Z_\theta +\delta Z_\mu+\delta Z_\phi).\\
\end{array}
\label{eqdeltaci}
\end{equation}

Taking into account eqs.~\eqref{zisvalues} and~\eqref{ordZ}, one concludes that the two equalities in eq.~\eqref{eqztheta} hold if, and only if,
\begin{equation}
\delta Z_{\theta}=0.
\label{vandelZ}
\end{equation}
This equation leads to the conclusion that $\theta^{\m\n}$ is not
renormalized at the one-loop level in the MS scheme of dimensional
regularization.

That the two equalities in eq.~\eqref{eqztheta} hold is a
necessary and sufficient condition for the gauge sector of our model --no matter fields in the external legs of the Green functions-- to be
multiplicatively renormalizable at one-loop and at first order in
$\theta^{\m\n}$. BRST invariance does not imply that eq.~\eqref{eqztheta} must
be verified, since in our case the most general BRST invariant contribution
involving only gauge fields reads up to first order in $h\theta^{\m\n}$:
\begin{equation}
\idx\; -\frac{1}{4}\,w_1\,f_{\m\n}f^{\m\n}\,+\,h\,\frac{e}{8}\,z_2\,t_1\,-\,h\,\frac{e}{2}\,
z_3\,t_2,
\label{3monomials}
 \end{equation}
Mark that the real numbers $w_1$, $z_2$ and $z_3$ are arbitrary.
Now, only if $z_2=z_3$, it is possible to renormalize the
$\theta^{\m\n}$ dependent part of the functional in the previous
equation by means of the renormalization in eq.~\eqref{mulren}. Of
course, we have shown by explicit computation that for our model
$z_2=z_3$. But there is more: we have obtained not only that
$z_2=z_3$, but that $z_2=z_3=w_1$. The latter train of equalities
has nothing to do with the the gauge sector of the model being
renormalizable at one loop, but with the fact that $\theta^{\m\n}$
is not renormalized at one-loop. We do not believe --following the
author of ref.~\cite{Wulkenhaar:2001sq}-- that this situation
--that $z_2=z_3=w_1$-- is an accident, but that it perhaps hints
at the existence of an as yet unknown symmetry  that mixes the
three monomials in eq.~\eqref{3monomials}. This symmetry must
depend on $\theta^{\m\n}$, for it must relate monomials with
different powers in $\theta^{\m\n}$. Notice that what we have
obtained is that the renormalizability of the gauge sector of the
model at one-loop and first order in $\theta^{\m\n}$ is governed
by the renormalization of the coupling constant $e$ --recall that
BRST invariance implies $\delta Z_a=-2\delta Z_e$.

Now, the matter sector of the model in the symmetric phase will be
multiplicatively renormalizable --i.e by means of the
renormalization transformations in eq.~\eqref{mulren}--  if, and
only if,  there exist $\delta \kappa_i$, $i=1,\dots,4$,
$Re\delta\kappa_5$ and $Im\delta\kappa_5$ such that the set of
eqs.~\eqref{eqdeltaci} holds for them. Taking into account  the
values of the $z_i$s on the r.h.s of eq.~\eqref{eqdeltaci} that
are given in eqs.~\eqref{zisvalues}, \eqref{deltais} and \eqref{delta8}, using the definitions of the $\delta C_i$s,
$i=1,\dots,9$, provided in eq.~\eqref{deltaCs}, and recalling that
the renormalized $C_i$s, $i=1,\dots,9$, are defined in terms of
the renormalized $\kappa_i$, $i=1,\dots, 5$, by the identities in
eq.~\eqref{coeffS1} and that  the values of the $\delta Z$s are those in
eqs.~\eqref{ordZ} and~\eqref{vandelZ}, one concludes, upon
substitution of the previous results, that there is a unique set
of parameters $\delta \kappa_i$, $i=1,\dots 5$, that solves the
system of equations constituted by the first five --$i=3,4,5,6$
and $7$-- equalities in eq.~\eqref{eqdeltaci}. This set of
parameters reads
\begin{equation}
  \begin{array}{l}
  \delta\kappa_1=\delta\text{Im}\ka_5-\frac{e^2}{32\pi^2\epsilon}(2\ka_1+\ka_2-2\text{Im}\ka_5+\ka_4)\\
\delta\ka_2=-\d\ka_4\\
\delta\ka_3=\frac{1}{192\pi^2\epsilon}[e^2(6+40 \ka_3)-\lambda(1+12\ka_3)]\\
\delta\text{Re}\ka_5=\frac{e^2}{128\pi^2\epsilon}(5-8\text{Re}\ka_5).
\end{array}
\label{soldeltaC}
\end{equation}
    And yet, the full system of equations has no solution for
$\mu^2\neq 0$, as the last equation --the equation with $z_9$ on the l.h.s-- is not satisfied by the  $\delta\kappa_i$s, $i=1,\dots,5$ in
eq.~\eqref{soldeltaC}. Indeed, upon substitution of those values in this last
equation one obtains the constraint $6e^2-\lambda=0$. Notice that this constraint is not even renormalization group invariant, so it cannot be imposed in a renormalization group invariant way, precluding the implementation of the reduction-of-the-couplings mechanism of ref.~\cite{Kubo:1988jc} to dispose of the unwanted
UV divergences. We thus conclude that the matter sector of the theory is not multiplicatively renormalizable if the scalar field is massive.
If $\mu^2=0$, the last equality of eq.~\eqref{eqdeltaci} need not be satisfied
since, now, terms of the type $\mu^2t_9$ occur neither in the classical action nor in $\Gamma^{(1)}[a_\mu,\phi,\phi^*]_{\rm pole}^{\rm
 one-loop}$ --see eq.~\eqref{wscoeff}. Unfortunately, the equation
$-z_8=\delta C_8+C_8(\delta Z_\theta +2\delta Z_\phi)$ is not
satisfied by the parameters given in eq.~\eqref{soldeltaC}, for
its substitution in the latter equation leads to the constraint
$204e^2-76e^2\lambda+15\lambda^2=0$. Remarkably, all dependence on
the arbitrary parameters of the Seiberg-Witten map disappears, but
the previous constraint is, of course, not valid for arbitrary $e$
and $\lambda$. The constraint is not even renormalization group
invariant. In summary, the matter sector of our model is not
multiplicatively renormalizable in the phase with no spontaneous
symmetry breaking whatever the value of the mass.

We shall next address the issue of the non-multiplicative
renormalizability of the model. We shall show that turning
non-multiplicative --but local at every order in $\theta^{\m\n}$--
the relationship between bare and renormalized fields will be of
no avail in making the model renormalizable. Let us assume that
the bare fields and renormalized fields are not related as in
eq.~\eqref{mulren}, but as follows
\begin{equation*}
a_{\mu}=a_\mu^{R}+\frac{1}{2}\,\delta Z_a \,a_\mu^{R}+ h \delta
Z_\mu[a_\m^{R},\phi^{R},\phi^{R\,*},\partial,\theta^{\m\n}];\quad\phi=\phi^{R}+\frac{1}{2}\delta
Z_{\phi}\,\phi^{R}+ h\delta
Z[a_\m^{R},\phi^{R},\phi^{R\,*},\partial_\m,\theta^{\m\n}],
\end{equation*}
where
\begin{equation}
 \begin{array}{l}
 \delta Z_\mu=x_1 {\theta_\m}^\a a_\a \phi^*\phi+x_2{\theta_\m}^\a a_\a (\phi\phi+\phi^*\phi^*) +ix_3{\theta_\m}^\a a_\a (\phi\phi-\phi^*\phi^*)+x_4{\theta_\m}^\a  (\Da\phi\phi+\Da\phi^*\phi^*)\\
\phantom{Z_\mu=}+ix_5{\theta_\m}^\a
(\Da\phi\phi-\Da\phi^*\phi^*)+x_6 {\theta_\m}^\a a_\a a_\r
a^\r+x_7{\theta_\m}^\a a_\a \partial _\r a^\r+x_8
{\theta_\m}^\a\Dr a_\a a^\r +x_9{\theta_\m}^\a\partial^2
a_\a\\[2pt]
\phantom{\delta Z_\mu=}+x_{10}{\theta_\m}^\a \Da a_\r a^\r
+x_{11} \mu^2{\theta_\mu}^\alpha \aa+\mu x_{12}{\theta_\mu}^\alpha \Da(\phi+\phi^*)+i\mu x_{13}{\theta_\mu}^\alpha \Da(\phi-\phi^*)\\
\phantom{\delta Z_\mu=}+\mu x_{14}{\theta_\mu}^\alpha \aa(\phi+\phi^*)+i\m x_{15}{\theta_\mu}^\alpha \aa(\phi-\phi^*)+x_{16}\theta^{\a\b}\Da \ab\am+x_{17}\theta^{\a\b}\Dm\aa\ab\\
\phantom{\delta Z_\mu=}+x_{18}\theta^{\a\b}\Da\am\ab,\\
 \delta Z=z_{1}\theta^{\a\b}\Da \ab\phi^*+z_2\theta^{\a\b}\aa\Db\phi+z_3\theta^{\a\b}\aa\Db\phi^*+\m
 z_{4}\theta^{\a\b}f_{\a\b},\\
 \end{array}
\label{nonmult}
\end{equation}
with real $x_i$s and  complex $z_{i}$s. The previous $\delta
Z_\mu$ and $\delta Z$ are the most general polynomials of mass
dimension one that are linear in $\theta^{\m\n}$ and do not
contain any contribution that can be removed by modifying the
value of the free parameters of the Seiberg-Witten map in
eq.~\eqref{SWmap}.

    $\delta Z_\mu$ and $\delta Z$ in eq.~\eqref{nonmult} yield the following
    sum of new counterterms
\begin{equation*}
 \begin{array}{l}
 S^{(1)\rm new}_{\rm ct}=\!\idx [\delta Z_\m \Dr f^{\r\m}-i \delta Z_\mu(D^\mu\phi^*\phi-\phi^*D^\m\phi)-
 \delta Z^*D^2\phi-\delta Z D^2\phi^*-\mu^2(\phi^*\delta Z+\phi \delta Z^*)\\
\phantom{S^{\rm new}_{\rm ct}=\idx [}-\frac{\lambda}{2}(\delta
Z^*\phi^*\phi^2+\delta Z\phi\phi^{*2})].
\end{array}
\end{equation*}
Now, $S^{(1)\rm new}_{\rm ct}$ must be invariant under the BRST
transformations $sa_\mu=\partial_\m c$,
$s\phi=iec\phi$,$s\phi^*=-iec\phi^*$. A lengthy computation shows
that $sS^{\rm new}_{\rm ct}=0$ if, and only if, $x_i=0$, $\forall
i$, and $z_i=0$, $\forall i$.

\section{The model in the phase with  spontaneously  broken symmetry}

In the case $\mu^2=-m^2<0,\lambda>0$, the classical
Poincar\'e-invariant vacuum of the theory with the action in
eq.~\eqref{Snc} is given by $\phi^*\phi=\frac{2m^2}{\lambda}$. To
perform perturbative calculations in the quantum theory we have to
expand the fields around a given vacuum configuration. We choose
the following parametrization:
\begin{equation*}
 \phi=\frac{1}{\sqrt{2}}(v+\phi_1+i\phi_2),
\end{equation*}
with $v=\sqrt{\frac{4m^2}{\lambda}}$ at the classical level and
with $\phi_1$ and $\phi_2$ being real fields that vanish in the
classical vacuum.

    Since we are interested in the renormalization properties of the model,  we shall consider the following family of
    $R_\xi$-gauges to quantize it:
\begin{equation}
    S_{\rm gf}=\idx\; \frac{1}{2}\xi b^2\, +\,b(\Dm a^\mu+\xi \rho \phi_2),\quad S_{\rm gh}=\idx\;\overline{c}(-\partial^2-\xi\rho
    e(v+\phi_1))c.
\label{Rxigauge}
\end{equation}
$b$ is an auxiliary real field and $c$ and $\bar{c}$ are the ghost
and anti-ghost fields, respectively. Recall that it is most useful
to choose $\rho=ev$ at the tree-level.

Now, up to first order in $h\theta^{\m\n}$, the action that we shall use to
carry out a path integral quantization of the theory reads
\begin{equation}
\begin{array}{l}
S_{\rm SSB}=
S^{(0)}[\phi=1/\sqrt{2}(v+\phi_1+i\phi_2),a_\mu,\mu^2=-m^2,\lambda]+
S_{\rm gf}+S_{\rm gh}\\
\phantom{S_{\rm SSB}=}+h S^{(1)}[\phi=1/\sqrt{2}(v+\phi_1+i\phi_2),a_\mu,\mu^2=-m^2,\lambda,\theta^{\m\n}],\\
\end{array}
\label{actionssb}
\end{equation}
where $S^{(0)}$ and $S^{(1)}$ have been defined in
eqs.~\eqref{Sord} and~\eqref{S1}, 
respectively, and $S_{\rm gf}$ and $S_{\rm gh}$ are given in
eq.~\eqref{Rxigauge}. Upon integrating over the auxiliary field
$b$, the previous action leads to the  set  of Feynman rules
depicted in figure 5 of the appendix. The following definitions
are needed to turn the Feynman rules into mathematical expressions
--notice that
$\tilde{\Gamma}^{t(i)\,\m_1,\mu_2,\dots,\mu_m}_{(m,n,p,q)}[{\rm
momenta}]$, $i=0,1$, denotes a tree-level vertex with $m$ fields
$a_\m$, $n$ fields $\phi_1$, $p$ fields $\phi_2$ and $q$ pairs,
$(c,\bar{c})$, of ghost-anti-ghost fields:

{\bf Propagators}
\begin{equation*}
\begin{array}{l}
 a_\m\longrightarrow G^{\m\n}(k)=\frac{-i}{k^2-(ev)^2+i\varepsilon}\Big[g^{\m\n}-(1-\xi)\frac{k^\mu k^\nu}{k^2-\xi(ev)^2}\Big]\\
  \phi_1\longrightarrow G_1(k)=\frac{i}{k^2-2m^2+i\varepsilon}\\
 \phi_2\longrightarrow G_2(k)=\frac{i}{k^2-\xi(ev)^2+i\varepsilon}\\
 c\longrightarrow G(k)=\frac{i}{k^2-\xi(ev)^2+i\varepsilon}\\
\end{array}
\end{equation*}\par
{\bf Ordinary vertices}
\begin{equation*}
\begin{array}{lll}
 \tilde{\Gamma}^{t(0)\,\m}_{(1,1,1,0)}[r;p;q]=e(p^\m-q^\m) &  \tilde{\Gamma}^{t(0)\,\m\n}_{(2,1,0,0)}[p,q;r]=2 i e^2 v g^{\m\n} &
\tilde{\Gamma}^{t(0)}_{(0,3,0,0)}[p,q,r]=-\frac{3}{2}i\lambda v \\
 \tilde{\Gamma}^{t(0)}_{(0,1,2,0)}[p;q,r]=-\frac{1}{2}i\lambda v & \tilde{\Gamma}^{t(0)}_{(0,1,0,2)}[p;q,r]=-i\xi e^2 v &
\tilde{\Gamma}^{t(0)\m\n}_{(2,2,0,0)}[p,q;r,s]=2ie^2g^{\m\n}\\
\tilde{\Gamma}^{t(0)\m\n}_{(2,0,2,0)}[p,q;r,s]=2ie^2g^{\m\n} & \tilde{\Gamma}^{t(0)}_{(0,4,0,0)}[p,q,r,s]=-\frac{3}{2}i\lambda & \tilde{\Gamma}^{t(0)}_{(0,0,4,0)}[p,q,r,s]=-\frac{3}{2}i\lambda\\
\tilde{\Gamma}^{t(0)}_{(0,2,2,0)}[p,q;r,s]=-\frac{1}{2}i\lambda
\end{array}
\end{equation*}
\par

{\bf Noncommutative vertices}

\begin{equation*}
 \begin{array}{l}
\tilde{\Gamma}^{t(1)\m}_{(1,1,0,0)}[r;q]=\frac{v}{2}(\Gamma^{t(1)\mu}_{(1,1)}[r;0,q]+\Gamma^{t(1)\mu}_{(1,1)}[r;-q,0])+\frac{v^3}{4}\Gamma^{t(1)\mu}_{(1,2)}[r;0,q;0,0]\\
\tilde{\Gamma}^{t(1)\m}_{(1,0,1,0)}[r;q]=\frac{-iv}{2}(\Gamma^{t(1)\mu}_{(1,1)}[r;-q,0]-\Gamma^{t(1)\mu}_{(1,1)}[r;0,q])\\
\tilde{\Gamma}^{t(1)\m\n}_{(2,0,0,0)}[r,s]=\frac{v^2}{2}\Gamma^{t(1)\mu\nu}_{(2,1)}[r,s;0,0]\\
\tilde{\Gamma}^{t(1)\m}_{(1,2,0,0)}[r;p,q]=\frac{1}{2}(\Gamma^{t(1)\mu}_{(1,1)}[r;-p,q]+\Gamma^{t(1)\mu}_{(1,1)}[r;-q,p])+\frac{3}{4}v^2\Gamma^{t(1)\mu}_{(1,2)}[r;0,p;0,q] \\
 \tilde{\Gamma}^{t(1)\m}_{(1,0,2,0)}[r;p,q]=\frac{1}{2}(\Gamma^{t(1)\mu}_{(1,1)}[r;-p,q]+\Gamma^{t(1)\mu}_{(1,1)}[r;-q,p])+\frac{1}{4}v^2\Gamma^{t(1)\mu}_{(1,2)}[r;0,p;0,q]\\
\tilde{\Gamma}^{t(1)\m}_{(1,1,1,0)}[r;q;p]=-\frac{i}{2}(\Gamma^{t(1)\m}_{(1,1)}[r;-p,q]-\Gamma^{t(1)\m}_{(1,1)}[r;-q,p])\\
\tilde{\Gamma}^{t(1)\m\n\e}_{(3,0,0,0)}[r,s,t]=\Gamma^{t(1)\m\n\e}_{(3,0)}[r,s,t]+\frac{v^2}{2}\Gamma^{t(1)\m\n\e}_{(3,1)}[r,s,t;0,0]\\
\tilde{\Gamma}^{t(1)\m\n}_{(2,1,0,0)}[r,s;q]=\frac{v}{2}(\Gamma^{t(1)\m\n}_{(2,1)}[r,s;0,q]+\Gamma^{t(1)\m\n}_{(2,1)}[r,s;-q,0])\\
\tilde{\Gamma}^{t(1)\m\n}_{(2,0,1,0)}[r,s;q]=\frac{-iv}{2}(\Gamma^{t(1)\m\n}_{(2,1)}[r,s;-q,0]-\Gamma^{t(1)\m\n}_{(2,1)}[r,s;0,q])\\
\tilde{\Gamma}^{t(1)\m\n}_{(2,2,0,0)}[r,s;p,q]=\frac{1}{2}(\Gamma^{t(1)\m\n}_{(2,1)}[r,s;-p,q]+\Gamma^{t(1)\m\n}_{(2,1)}[r,s;-q,p])\\
\tilde{\Gamma}^{t(1)\m\n}_{(2,0,2,0)}[r,s;p,q]=\frac{1}{2}(\Gamma^{t(1)\m\n}_{(2,1)}[r,s;-p,q]+\Gamma^{t(1)\m\n}_{(2,1)}[r,s;-q,p])\\
\tilde{\Gamma}^{t(1)\m\n}_{(2,1,1,0)}[r,s;q,p]=\frac{-i}{2}(\Gamma^{t(1)\m\n}_{(2,1)}[r,s;-p,q]-\Gamma^{t(1)\m\n}_{(2,1)}[r,s;-q,p])\\
\tilde{\Gamma}^{t(1)\m\n\r}_{(3,1,0,0)}[r,s,t;q]=v\Gamma^{t(1)\m\n\r}_{(3,1)}[r,s,t;0,q]\\
\tilde{\Gamma}^{t(1)\m}_{(1,3,0,0)}[s;p,q,r]=\frac{3v}{2}\Gamma^{t(1)\m}_{(1,2)}[s;0,p;-q,r]\\
\tilde{\Gamma}^{t(1)\m}_{(1,1,2,0)}[s;p;q,r]=\frac{v}{2}\Gamma^{t(1)\m}_{(1,2)}[s;0,p;-q,r]\\
\tilde{\Gamma}^{t(1)\m\n\r}_{(3,2,0,0)}[r,s,t;p,q]=\Gamma^{t(1)\m\n\r}_{(3,1)}[r,s,t;-p,q]\\
\tilde{\Gamma}^{t(1)\m\n\r}_{(3,0,2,0)}[r,s,t;p,q]=\Gamma^{t(1)\m\n\r}_{(3,1)}[r,s,t;-p,q]\\
\tilde{\Gamma}^{t(1)\m}_{(1,4,0,0)}[t;p,q,r,s]=\frac{3}{2}\Gamma^{t(1)\m}_{(1,2)}[t;-p,q;-r,s]\\
\tilde{\Gamma}^{t(1)\m}_{(1,0,4,0)}[t;p,q,r,s]=\frac{3}{2}\Gamma^{t(1)\m}_{(1,2)}[t;-p,q;-r,s]\\
 \tilde{\Gamma}^{t(1)\m}_{(1,2,2,0)}[t;p,q;r,s]=\frac{1}{2}\Gamma^{t(1)\m}_{(1,2)}[t;-p,q;-r,s]\\
 \end{array}
\label{feynmanssb}
\end{equation*}
with the $\Gamma^t$s as given in eq.~\eqref{feynmannossb}, but evaluated at $\mu^2=-m^2$. All momenta are taken as positive when coming out of the vertex.

Before discussing the renormalizablity at first order in
$\theta^{\m\n}$ of the model in the phase with spontaneous
symmetry breaking, we shall just remark the obvious fact  that the
one-loop UV divergent contributions that do not depend on
$\theta^{\m\n}$ --i.e., the one-loop UV divergent contributions of
the ordinary model-- can be multiplicatively 
renormalized --see refs.~\cite{Appelquist:1973ms, Vargas:1990nw, Petriello:2001mp} for further details-- by expressing the  bare fields and parameters--denoted by the superscript $0$-- in terms of the renormalized fields and 
parameters --labelled with the superscript ``$R$''-- as follows:
\begin{equation}
 \begin{array}{lll}
 a_\m^0=Z_a^{1/2}a^R_\m   &   \phi_1^0=Z_{\phi_1}^{1/2}\phi_1^R & \phi_2^0=Z_{\phi_2}^{1/2}\phi_2^R\\
v_0=Z_{\phi_1}^{1/2}(v^R+\delta v) & e^0=Z_e e^R &  m^0=Z_m^{1/2}m^R\\
   \lambda^0=Z_\lambda \lambda^R.\\
\end{array}
\label{Zssb}
\end{equation}
In the MS scheme of dimensional regularization --recall that $D=4+2\epsilon$--
 one has that $Z_{\phi_1}=Z_{\phi_2}=Z_\phi$, with $Z_\phi$ given in
eq.~\eqref{ordZ}, and that $Z_a,\,Z_e,\,Z_m=Z_\mu,\, Z_\la$ take
the same  values as in the phase with no spontaneous symmetry
breaking -see eq.~\eqref{ordZ}--, if
\begin{equation*}
  \frac{\delta v}{v^R}=\frac{-e^2\xi}{16\pi^2\epsilon}. 
\end{equation*}


\subsection{\bf One-loop renormalizability of the gauge sector}

In dimensional regularization, the pole part of  any UV divergent one-loop Feynman integral, $I_F$, is a polynomial on the external
momenta of the integral and the masses of the free internal propagators, if it  is besides IR finite by power counting at
non-exceptional momenta. Further, if the Feynman integral, say $I_F(m=0)$,  that is obtained from $I_F$ by setting to zero all
the masses in the denominators is still IR finite by power counting at non-exceptional momenta, there happens that the
pole part of $I_F$ that does not depend on the masses is  given by the pole part of the integral $I_F(m=0)$.

For the remaining of this subsection, to render both the
computations and the subsequent analysis as simple as possible, we
shall send to zero the gauge parameter, $\xi$, that occurs in the
Feynman rules of the model --these rules are given in figure 5 of
the appendix. This way the interaction vertex involving the ghost
fields vanishes. Let $\Gamma^{(1)}[a_\mu]^{\rm
SSB,\,one-loop}_{\rm pole}$ denote the one-loop pole part of the
1PI functional of the gauge sector of the model at first order in
$\theta^{\m\n}$ --by definition $\Gamma^{(1)}[a_\mu]^{\rm
SSB,\,one-loop}_{\rm pole}$ only depends on $a_\m$. Taking into
account the arguments presented in the previous paragraph, one
concludes that the contributions to $\Gamma^{(1)}[a_\mu]^{\rm
SSB,\,one-loop}_{\rm pole}$ that do not depend on any dimensionful parameter 
--that we shall denote with $M$-- are equal to those in the massless theory, 
which were obtained  in the previous section:
\begin{equation}
\begin{array}{l}
\Gamma^{(1)}[a_\mu]^{\rm SSB,\;one-loop}_{\rm pole}=
\Gamma^{(1)}[a_\mu]^{M-{\rm independent}}_{pole}+\Gamma^{(1)}[a_\mu]^{M-{\rm dependent}}_{pole},\\
\Gamma^{(1)}[a_\mu]^{M-{\rm independent}}_{pole}=\idx\;\big(\frac{e}{8}\,z_1\,t_1-\frac{e}{2}\,z_2\,t_2\,\big).\\
\end{array}
\label{polegauge}
\end{equation}
$t_1$ and $t_2$ were defined in eq.~\eqref{tbasis}, and $z_1$ and $z_2$ were given in eq.~\eqref{zisvalues} --see also eq.~\eqref{deltais}.
$\Gamma^{(1)}[a_\mu]^{M-{\rm dependent}}_{pole}$ --the $M-$dependent
contribution to $\Gamma^{(1)}[a_\mu]^{\rm SSB,\;one-loop}_{\rm pole}$-- can be obtained from  the pole of the $M$-dependent part of the one-loop 1PI diagrams contributing to $<0|T\{a_\m(x) a_\n(y)\}|0>$ and $<0|T\{a_\m(x) a_\n(y) a_\rho(z)\}|0>$. The topologically inequivalent diagrams that contribute at first first order in $\theta^{\m\n}$ are given in figures 6 and 7 of the appendix. It turns out that
\begin{equation}
\begin{array}{l}
\Gamma^{(1)}[a_\mu]^{M-{\rm dependent}}_{pole}=
\frac{(e v)^2}{2}\,\theta^{\a\b}\idx\;\Big(i\,\Delta_4^{(\xi=0)}\,a_\rho \partial^{\rho}f_{\a\b}+
i\,\Delta_5^{(\xi=0)}\,a_\a \partial_{\rho}{f_{\b}}^{\rho}+\\
\phantom{\Gamma^{(1)}[a_\mu]^{M-{\rm dependent}}_{pole}=
(e v)^2\,\theta^{\a\b}\idx\;\big(i\,\Delta_4^{(\xi=0)}\,a_\rho +}
e\,\Delta_6^{(\xi=0)}\,a_\rho a^{\rho}f_{\a\b}+
e\,\Delta_7^{(\xi=0)}\,a_\a a_{\rho}{f_{\b}}^{\rho}\Big),\\
\end{array}
\label{vpolegauge}
\end{equation}
where $\Delta_4^{(\xi=0)}$, $\Delta_5^{(\xi=0)}$, $\Delta_6^{(\xi=0)}$ and $\Delta_7^{(\xi=0)}$ are obtained by substituting $\xi=0$ in $\Delta_4$, $\Delta_5$, $\Delta_6$ and $\Delta_7$, as given in eq.~\eqref{deltais}, respectively.

Let us now show that the UV divergences in eqs.~\eqref{polegauge}
and eq.~\eqref{vpolegauge} can be removed by renormalizing the
parameters and fields as in eq.~\eqref{Zssb}, if we also introduce
the following renormalization of the parameters $\kappa_i$, $i=1,\dots,5$,
of the Seiberg-Witten map in eq.~\eqref{SWmap}:
\begin{equation}
\kappa_i^0=\kappa_i^R+\delta\kappa_i,\,i=1,2,3,4,\quad \text{Re}\ka_5^0=\text{Re}\ka_5^R+\delta\text{Re}\ka_5^R,\quad
\text{Im}\ka_5^0=\text{Im}\ka_5^R+\delta\text{Im}\ka_5^R.
\label{kapparen}
\end{equation}
For completeness one  should also include the following renormalization of
$\theta^{\m\n}$: $\theta^{0\,\m\n}=\Z_\theta \theta^{R\,\m\n}$, but as we shall see the renormalization of the gauge sector implies $Z_\theta=1$ at the order at which we are working.

The substitution of the definitions in eqs.~\eqref{Zssb} and~\eqref{kapparen}
in the action in eq.~\eqref{actionssb} yields the following $\theta^{\m\n}-$dependent counterterms involving only gauge fields:
\begin{equation*}
\begin{array}{c}
S^{(1)}_{\rm ct}[a]=\idx\, \big[\,\frac{e}{8}\, (\delta Z_\theta +\delta Z_a) t_1-\frac{e}{2} (\delta Z_\theta+\delta Z_a)  t_2\,\big]\,+\\
\frac{(ev)^2}{2}\,\theta^{\a\b}\idx\,\big\{
i\,[\delta C_4+C_4(\delta Z_\theta +\delta Z_\phi)]\,a_\rho \partial^{\rho}f_{\a\b}\,+
i\,[\delta C_5+C_5(\delta Z_\theta +\delta Z_\phi)]\,a_\a \partial_{\rho}{f_{\b}}^{\rho}\big\}+\\
\frac{(ev)^2}{2}\,\theta^{\a\b}\idx\,\big\{e\,[\delta C_6+C_6(\delta Z_\theta +\delta Z_\phi)]\, a_\rho a^{\rho}f_{\a\b}\,+\,
e\,[\delta C_7+C_7(\delta Z_\theta +\delta Z_\phi)]\,
a_\a a_{\rho}{f_{\b}}^{\rho}\big\},\\
\end{array}
\end{equation*}
where $\delta C_4$, $\delta C_5$, $\delta C_6$ and $\delta C_7$ were defined in eq.~\eqref{deltaCs}. In obtaining
$S^{(1)}_{\rm ct}[a_\m]$ above, we have used the results: $\delta Z_a=-2\,\delta Z_e,\,\delta v^{(\xi=0)}=0$.

It is plain that $S^{(1)}_{\rm ct}[a_\m]$ defined in the MS scheme
will cancel $\Gamma^{(1)}[a_\mu]^{\rm SSB,\;one-loop}_{\rm pole}$
given by eqs.~\eqref{polegauge} and~\eqref{vpolegauge} if, and
only if,
\begin{equation*}
\begin{array}{c}
 \frac{e^2}{48\pi^2\epsilon}=\delta Z_\theta +\delta Z_a,\\
-2\Delta_4^{(\xi=0)}+\Delta_5^{(\xi=0)}=2\delta C_4-\delta C_5+(2C_4-C_5)(\delta Z_\theta +\delta Z_\phi),\\
-\Delta_i^{(\xi=0)}\,=\,\delta C_i\,+\, C_i(\delta Z_\theta +\delta Z_\phi),\quad i=6,7.\\
\end{array}
\end{equation*}
The previous set of equations is a subset of the set of equalities
constituted by eq.~\eqref{eqztheta} and the first five equalities
in eq~\eqref{eqdeltaci} evaluated at $\xi=0$. Hence, taking into
account that $\delta Z_a$ and $\delta Z_{\phi}$ have the same
value --given in eq.~\eqref{ordZ}-- as in the phase with unbroken
symmetry but with the choice $\xi=0$, one concludes first that
$\delta Z_\theta=0$ and second that by choosing $\delta \kappa_i$,
$i=1,\dots, 5$, as in eq.~\eqref{soldeltaC} --i.e., as in the
symmetric phase-- we will  be able to remove the UV divergences of
the gauge sector at one-loop and at first order in
$\theta^{\m\n}$.

Let us show next that the one-loop  renormalizability of the gauge sector of
the model in the phase with spontaneous symmetry breaking that we have just
discussed is a consequence of the two facts: {\it i)} that the $U(1)$ symmetry is broken spontaneously so that the action in eq.~\eqref{actionssb} is invariant under the following BRST transformations
\begin{equation*}
s a_\mu=\partial_{\mu}c,\quad s\phi_1=-ec\,\phi_2,\quad s\phi_2=ec\,(\phi_1\,+\,v),\quad sc=0,\quad s\bar{c}=b,\quad sb=0
\end{equation*}
and {\it ii)} that the pole part of the one-loop 1PI functional
that does not depend on $v$ is the same as in the massless model.
To use as simple as possible  linearized Slavnov-Taylor equations,
we shall still keep the gauge-fixing parameter $\xi$ equal to 0.
For this value of the gauge-fixing parameter the ghost and
anti-ghost fields decouple and, hence, they do not contribute to
the dimensionally regularized one-loop 1PI functional,
$\Gamma_{\rm SSB}$, obtained from our Feynman rules in figure 5 of
the appendix. Since the gauge-fixing equation
\begin{equation*}
\frac{\delta\Gamma_{\rm SSB}}{\delta b}=\xi\,b+\Dm a^\mu +\xi\rho\phi_2
\end{equation*}
holds for the dimensionally regularized 1PI functional
$\Gamma_{\rm SSB}$ obtained from $S_{\rm SSB}$ in
eq.~\eqref{actionssb}, it turns out that in the gauge $\xi=0$ the
BRST invariance of the model implies that the one-loop
contribution, $\Gamma_{\rm SSB}^{\rm one-loop}$, to $\Gamma_{\rm
SSB}$ is a function of $\phi=\frac{1}{\sqrt{2}}(v+\phi_1+i\phi_2)$
and $\phi^{*}$ that must satisfy the following linearized
Slavnov-Taylor equation
\begin{equation}
\iDx\, sa_\m(x)\;\frac{\delta\Gamma_{\rm SSB}^{\rm one-loop}}{\delta a_\m(x)}\,+\,
s\phi(x)\,\frac{\delta\Gamma_{\rm SSB}^{\rm one-loop}}{\delta \phi(x)}+
s\phi^{*}(x)\,\frac{\delta\Gamma_{\rm SSB}^{\rm one-loop}}{\delta \phi^{*}(x)}\,=\,0,
\label{stssb}
\end{equation}
where $s\phi=iec\phi$ and $s\phi^{*}=-iec\phi^{*}$.
eq.~\eqref{stssb} leads to the conclusion that when $\xi=0$ the
pole part of the one-loop 1PI functional  $\Gamma_{\rm SSB}^{\rm
one-loop}$ is given by the most general  gauge invariant local
polynomial which is a functional of  $a_\m$,
$\phi=\frac{1}{\sqrt{2}}(v+\phi_1+i\phi_2)$ and $\phi^{*}$ --it
must then be a local polynomial of $f_{\m\n}$, $\phi$ and $\phi^*$
and their gauge covariant derivatives. This result and the
analysis carried out in the first paragraph of this subsection
implies that for $\xi=0$ the pole contribution to $\Gamma_{\rm
SSB}^{\rm one-loop}$ that is linear in $\theta^{\m\n}$, say
$\Gamma_{\rm SSB}^{(1)\,{\rm one-loop}}$, reads
\begin{equation}
\Gamma_{\rm SSB}^{(1)\,{\rm one-loop}}=\idx\;\frac{e}{8}\,z_1^{(\xi=0)}\,t_1-\frac{e}{2}\,z_2^{(\xi=0)}\,t_2+\sum_{i=3}^8 e z_i^{(\xi=0)}\,t_i+e
r_9(m,v) t_9,
\label{predict}
\end{equation}
where $z_i^{(\xi=0)}$, $i=1,\dots,8$, are given  by the
corresponding $z_i$ in eq.~\eqref{zisvalues}, upon substituting
$\xi=0$, and $t_i$, $i=1,\dots,9$, are defined as in
eq.~\eqref{tbasis} but, now, with
$\phi=\frac{1}{\sqrt{2}}(v+\phi_1+i\phi_2)$. We have thus shown
that, for $\xi=0$, $\Gamma_{\rm SSB}^{(1)\,{\rm one-loop}}$ is a
linear combination of the basis of gauge invariant polynomials
given in eq.~\eqref{tbasis} with coefficients such that, when
$m$ and $v\rightarrow 0$, one recovers the corresponding object for the
massless Higgs-Kibble model at $\xi=0$. Finally,
eq.~\eqref{predict} leads to $\Gamma^{(1)}[a_\mu]^{\rm
SSB,\;one-loop}_{\rm pole}$ as given by eqs.~\eqref{polegauge}
and~\eqref{vpolegauge} upon imposing the condition $\xi=0$.

\subsection{Non-renormalizability of the matter sector}

Recall that we are in the phase with spontaneously broken gauge symmetry. 
Let $\Gamma^{M-{\rm independent}}_{\rm pole}[a_\mu,\phi_1,
\phi_2]$ denote the one-loop pole part of the  1PI functional of the model 
that does not depend on any dimensionful parameter $M$ for arbitrary $\xi$. 
Taking advantage of the discussion carried out in the first paragraph of 
the previous subsection, one concludes that $\Gamma^{M-{\rm independent}}_{\rm
pole}[a_\mu,\phi_1, \phi_2]$ is equal to the corresponding object
computed in the massless model. We have shown in the previous
section --section 3-- that there is no local way of renormalizing
the fields and parameters of the model that removes the UV
divergences of the matter sector of the massless model. Hence, in
the phase with spontaneous symmetry breaking, there is also no
local way of renormalizing the fields and parameters of the field
theory that substracts the $M-$independent UV divergent
contributions occurring at the one-loop level in the 1PI functional of
the matter sector of the model.

\newpage

\section{Summary and conclusions}

In this paper we have shown that the noncommutative $U(1)$
Higgs-Kibble model formulated within the enveloping-algebra
formalism of refs.\cite{Madore:2000en,Jurco:2000ja}
and~\cite{Jurco:2001rq} is non-renormalizable in perturbation
theory in the phase with unbroken gauge symmetry, whatever the
value of the mass of the complex scalar field. We have also shown
that the same result holds when the model is in the phase with
spontaneous symmetry breaking. However, the gauge sector of the
model is one-loop renormalizable at first order in $\theta^{\m\n}$
whatever the phase we look at. This is quite surprising --although
in keeping with the results obtained in
refs.\cite{Wulkenhaar:2001sq} and~\cite{Buric:2005xe} for other
models-- since gauge symmetry -either noncommutative or ordinary--
and power counting do not imply it --see discussion in the
paragraph below eq.~\eqref{vandelZ}. This renormalizability of the
gauge sector of the model appears even more surprising if we take
into account that the matter sector is non-renormalizable and that
all the one-loop  UV divergent diagrams that contribute to the gauge
sector in the phase with unbroken gauge symmetry --see figure 2--
have only scalar particles propagating along the loop. The
question thus arises as to whether the renormalizability of the
gauge sector of all the models studied so far, hints at the
existence of an as yet unveiled new symmetry of these gauge models
so that the part  of the 1PI functional that only depends on the gauge
fields is constrained by it. The existence of such a symmetry
will be of paramount importance in modifying the matter sector so
that it becomes renormalizable. Finally, the results presented in this paper
make us confident that all the one-loop UV divergent contributions to the 
gauge sector of the noncommutative standard model coming from the matter sector of the model are renormalizable, at least at first order in $\theta^{\m\n}$. Hence, phenomenological results such as those obtained in ref.~\cite{Buric:2006nr} are robust due to the one-loop renormalizability of the gauge sector. 

\section{Acknowledgements}

This work has been financially suported in part by MEC through grant
FIS2005-02309. The work of C. Tamarit has also received financial support
from MEC through FPU grant AP2003-4034. C. Tamarit should like to thank
Dr. Chong-Sun Chu for valuable conversations and the Department of
Mathematical Sciences of the University of Durham, United Kingdom, where part
of this work was carried out, for its kind hospitality.

\section{Appendix. Feynman rules and Feynman diagrams with a noncommutative vertex}

In this appendix we collect the figures with the Feynman rules and
1PI Feynman diagrams that are referred to in the main text of the
paper. In figure 1, the Feynman rules of our noncommutative Higgs-Kibble model
in the phase with unbroken gauge symmetry are given. The topologically inequivalent Feynman diagrams contributing to $\Gamma^{(1)\m\n\r}_{(3,0)}[p_1,p_2,-p_1-p_2]_{\rm pole}^{\rm
one-loop}$, $\Gamma^{(1)\m}_{(1,1)}[p_1-p_2;p_1,p_2]_{\rm pole}^{\rm
one-loop}$ and $\Gamma^{(1)\m}_{(1,2)}[t;p,q;r,s]_{\rm
pole}^{\rm one-loop}$ are depicted in figures 2, 3 and 4. The Feynman rules
of our non-commutative  Higgs-Kibble model in the phase with spontaneous symmetry breaking are drawn in figure 5. Finally, in figures 6 and 7, the topologically inequivalent Feynman diagrams contributing to the pole part of the $M$-dependent part of the 1PI functions of the gauge field are shown.
\newpage
\begin{minipage}{0.96\textwidth}
\centering
\begin{minipage}[c]{0.19\textwidth}\centering
\epsfig{file=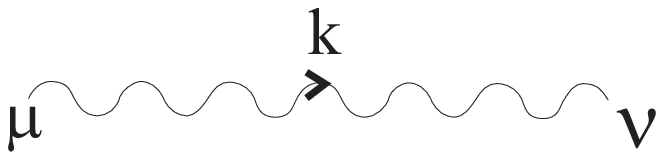,height=0.6cm,width=3cm}
\end{minipage}%
\begin{minipage}[c]{0.17\textwidth}\centering
\small
\flushleft
\begin{equation*}
    \leftrightarrow  G^{\m\n}[k]
\end{equation*}
\end{minipage}%
\begin{minipage}[c]{0.19\textwidth}\centering
\epsfig{file=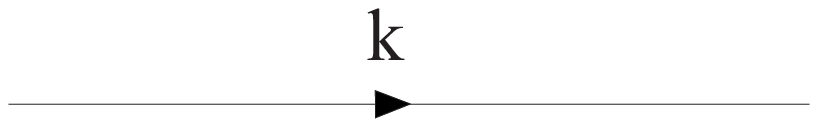,height=0.6cm,width=3cm}
\end{minipage}%
\begin{minipage}[c]{0.17\textwidth}
\small
\begin{equation*}
    \leftrightarrow  G[k]
\end{equation*}
\end{minipage}%
\par\centering
\begin{minipage}[c]{0.13\textwidth}
\epsfig{file=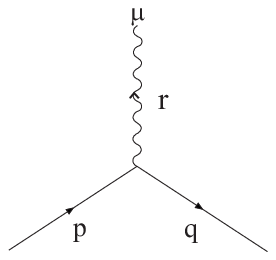,height=2cm}
\end{minipage}%
\begin{minipage}[c]{0.17\textwidth}
\small
\begin{equation*}
    \leftrightarrow\Gamma^{t(0)\,\m}_{(1,1)}[r;p,q]
\end{equation*}
\end{minipage}%
\begin{minipage}[c]{0.15\textwidth}\centering
\epsfig{file=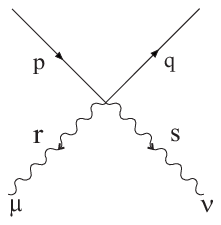,height=2cm}
\end{minipage}%
\begin{minipage}[c]{0.20\textwidth}
\small
\begin{equation*}
        \leftrightarrow \Gamma^{t(0)\,\m\n}_{(2,1)}[r,s;p,q]
     \end{equation*}
\end{minipage}%
\begin{minipage}[c]{0.13\textwidth}\centering
\epsfig{file=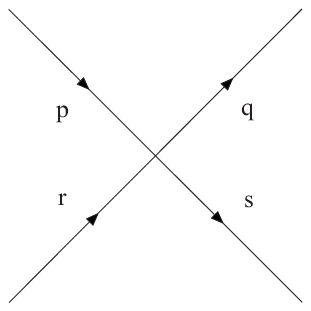,height=2cm}
\end{minipage}%
\begin{minipage}[c]{0.20\textwidth}
\small
 \begin{equation*}
\leftrightarrow\Gamma^{t(0)}_{(0,2)}[r,s;p,q]
\end{equation*}
\end{minipage}
\begin{minipage}[c]{0.13\textwidth}\centering
\epsfig{file=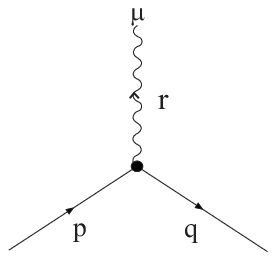,height=2cm}
\end{minipage}%
\begin{minipage}[c]{0.17\textwidth}
\small
 \begin{equation*}
\leftrightarrow \Gamma^{t(1)\,\m}_{(1,1)}[r;p,q]
\end{equation*}
\end{minipage}
\begin{minipage}[c]{0.13\textwidth}\centering
\epsfig{file=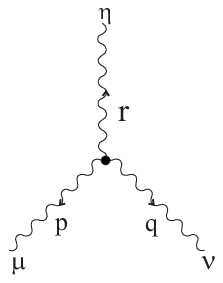,height=2cm}
\end{minipage}%
\begin{minipage}[c]{0.18\textwidth}
\small
 \begin{equation*}
\leftrightarrow \Gamma^{t(1)\,\m\n\e}_{(3,0)}[p,q,r]
\end{equation*}
\end{minipage}
\begin{minipage}[c]{0.15\textwidth}\centering
\epsfig{file=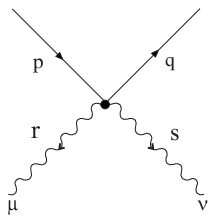,height=2cm}
\end{minipage}%
\begin{minipage}[c]{0.20\textwidth}
\small
 \begin{equation*}
\leftrightarrow \Gamma^{t(1)\,\m\n}_{(2,1)}[r,s;p,q]
\end{equation*}
\end{minipage}\par\centering
\begin{minipage}[c]{0.13\textwidth}\centering
\epsfig{file=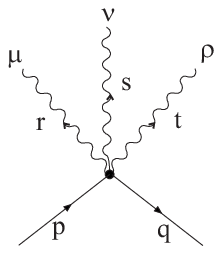,height=2cm}
\end{minipage}%
\begin{minipage}[c]{0.25\textwidth}
\small
 \begin{equation*}
 \begin{array}{l}
\leftrightarrow \Gamma^{t(1)\,\m\n\r}_{(3,1)}[r,s,t;p,q]
\end{array}
\end{equation*}
\end{minipage}
\begin{minipage}[c]{0.13\textwidth}\centering
\epsfig{file=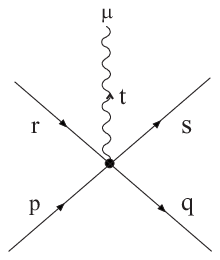,height=2cm}
\end{minipage}%
\begin{minipage}[c]{0.22\textwidth}
\small
 \begin{equation*}
 \begin{array}{l}
\leftrightarrow \Gamma^{t(1)\,\m}_{(1,2)}[t;p,q;r,s]
\end{array}
\end{equation*}
\end{minipage}
\\[13pt]
 \renewcommand{\figurename}{Fig.}
 \renewcommand{\captionlabeldelim}{.}
\figcaption{Feynman rules for the phase with unbroken symmetry.}
\end{minipage}
\\[\intextsep]
\\[\intextsep]
\begin{minipage}{0.96\textwidth}
\centering
\epsfig{file=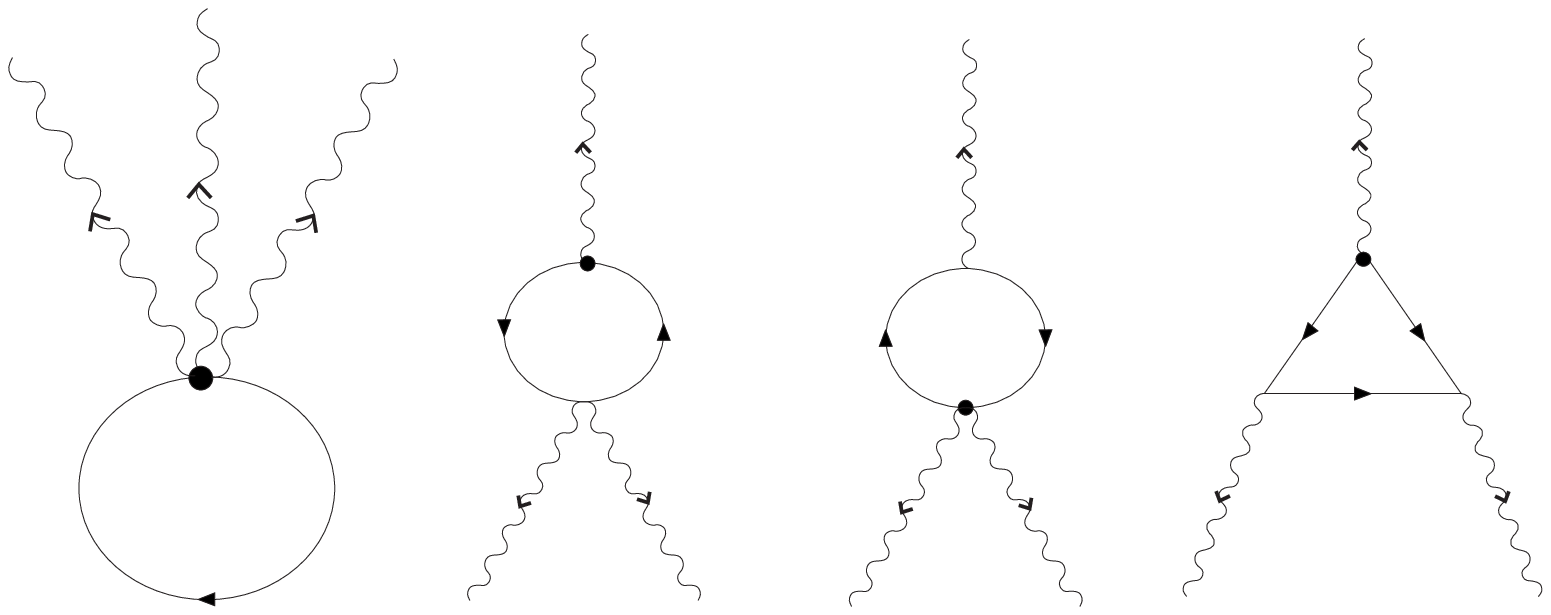,height=3cm}
\\[7pt]
 \renewcommand{\figurename}{Fig.}
 \renewcommand{\captionlabeldelim}{.}
\figcaption{Topologically inequivalent diagrams contributing to
$\Gamma^{(1)\m\n\r}_{(3,0)}[p_1,p_2,-p_1-p_2]_{\rm pole}^{\rm
one-loop}$.}
\end{minipage}
\\[\intextsep]
\\[\intextsep]
\begin{minipage}{0.96\textwidth}
\centering
\epsfig{file=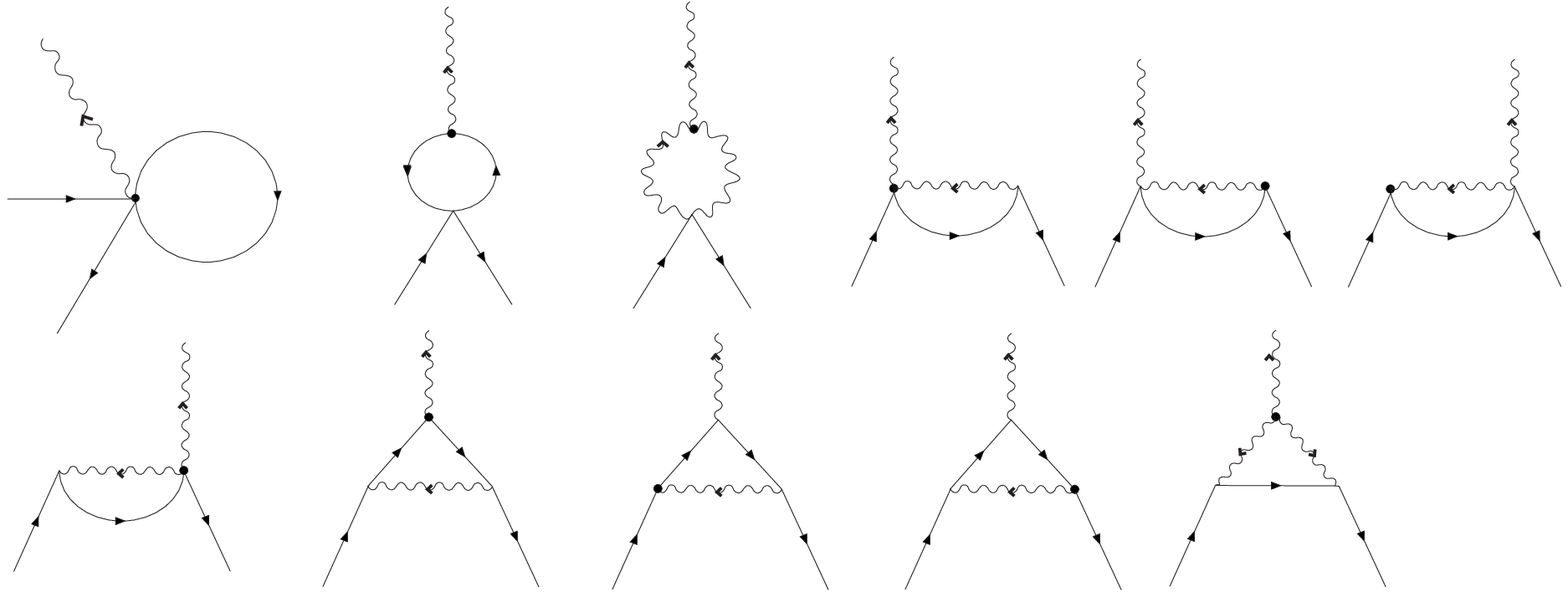,height=5cm}
\\[7pt]
 \renewcommand{\figurename}{Fig.}
 \renewcommand{\captionlabeldelim}{.}
\figcaption{Topologically inequivalent diagrams contributing to
$\Gamma^{(1)\m}_{(1,1)}[p_1-p_2;p_1,p_2]_{\rm pole}^{\rm
one-loop}$.}
\end{minipage}
\\[\intextsep]
\begin{minipage}{\textwidth}
\centering
\epsfig{file=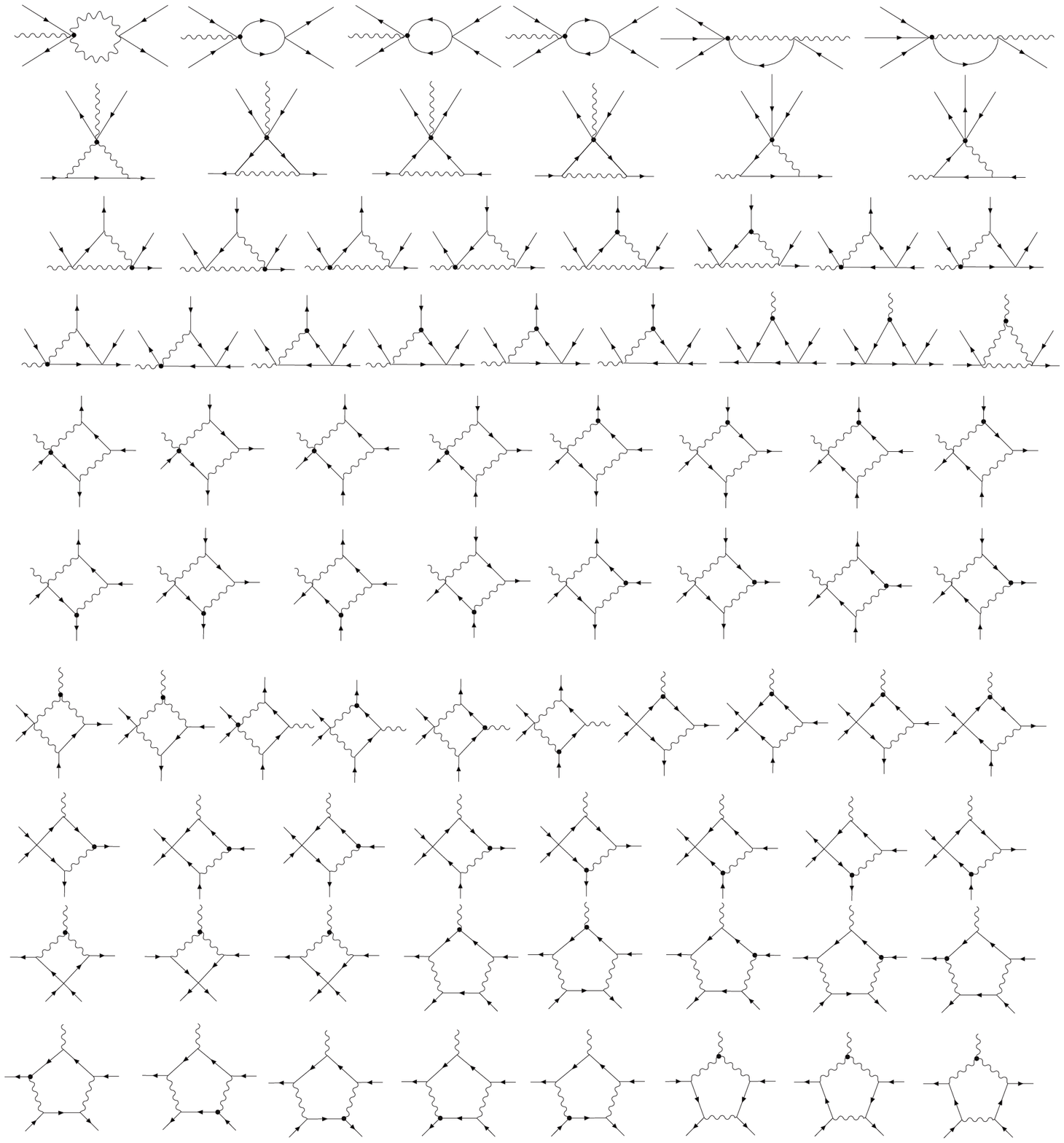,width=16.8cm}
\\[13pt]
 \renewcommand{\figurename}{Fig.}
 \renewcommand{\captionlabeldelim}{.}
\figcaption{Topologically inequivalent diagrams contributing to
$\Gamma^{(1)\m}_{(1,2)}[t;p,q;r,s]_{\rm
pole}^{\rm one-loop}$.}
\end{minipage}
\begin{minipage}{\textwidth}
\centering
\begin{minipage}[c]{0.27\textwidth}\centering
\epsfig{file=foton.eps,height=0.6cm,width=3cm}
\end{minipage}%
\begin{minipage}[c]{0.17\textwidth}
\small
\begin{equation*}
    \leftrightarrow  G^{\m\n}[k]
\end{equation*}
\end{minipage}
\begin{minipage}[c]{0.27\textwidth}\centering
\epsfig{file=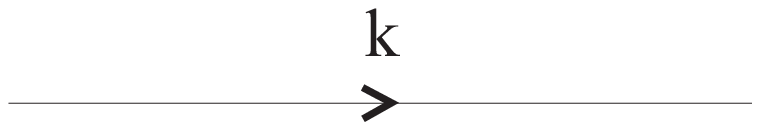,height=0.6cm,width=3cm}
\end{minipage}%
\begin{minipage}[c]{0.17\textwidth}
\small
\begin{equation*}
    \leftrightarrow  G_1[k]
\end{equation*}
\end{minipage}\par\centering
\begin{minipage}[c]{0.27\textwidth}\centering
\epsfig{file=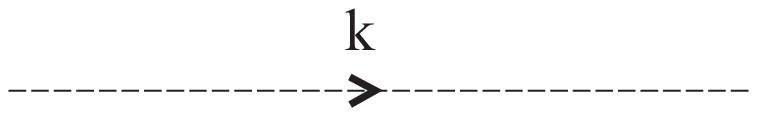,height=0.6cm,width=3cm}
\end{minipage}%
\begin{minipage}[c]{0.17\textwidth}
\small
\begin{equation*}
    \leftrightarrow  G_2[k]
\end{equation*}
\end{minipage}
\begin{minipage}[c]{0.27\textwidth}\centering
\epsfig{file=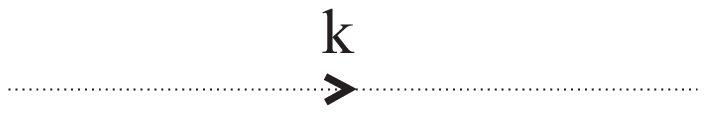,height=0.6cm,width=3cm}
\end{minipage}%
\begin{minipage}[c]{0.17\textwidth}
\small
\begin{equation*}
    \leftrightarrow  G[k]
\end{equation*}
\end{minipage}\par\centering
\begin{minipage}[c]{0.13\textwidth}
\epsfig{file=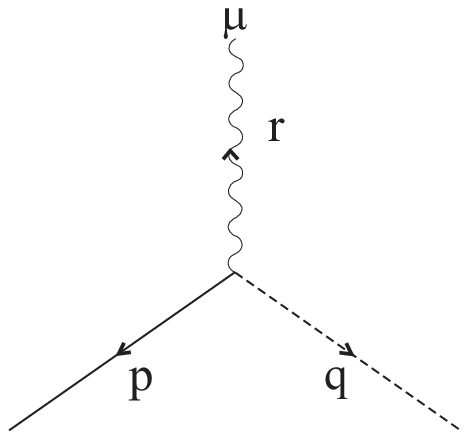,height=2cm}
\end{minipage}
\begin{minipage}[c]{0.18\textwidth}
\small
\begin{equation*}
    \leftrightarrow  \tilde{\Gamma}^{t(0)\,\m}_{(1,1,1,0)}[r;p;q]
\end{equation*}
\end{minipage}%
\begin{minipage}[c]{0.13\textwidth}
\epsfig{file=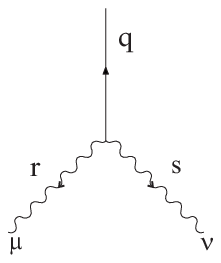,height=2cm}
\end{minipage}%
\begin{minipage}[c]{0.175\textwidth}
\small
\begin{equation*}
    \leftrightarrow  \tilde{\Gamma}^{t(0)\,\m\n}_{(2,1,0,0)}[r,s;q]
\end{equation*}
\end{minipage}%
\begin{minipage}[c]{0.13\textwidth}
\epsfig{file=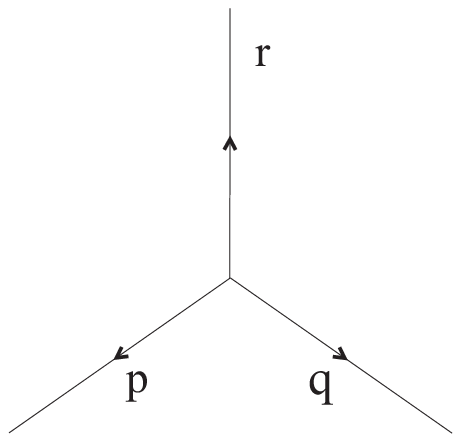,height=2cm}
\end{minipage}%
\begin{minipage}[c]{0.175\textwidth}
\small
\begin{equation*}
    \leftrightarrow  \tilde{\Gamma}^{t(0)}_{(0,3,0,0)}[p,q,r]
\end{equation*}
\end{minipage}\\
\begin{minipage}[c]{0.13\textwidth}
\epsfig{file=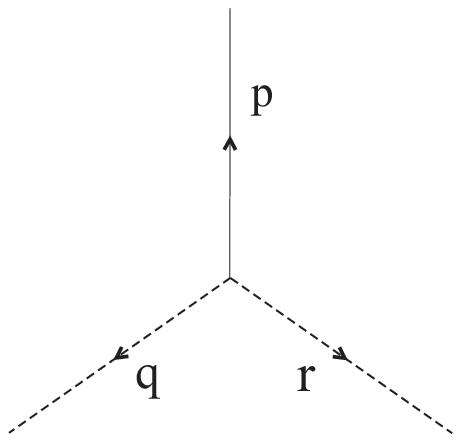,height=2cm}
\end{minipage}%
\begin{minipage}[c]{0.175\textwidth}
\small
\begin{equation*}
    \leftrightarrow  \tilde{\Gamma}^{t(0)}_{(0,1,2,0)}[p;q,r]
\end{equation*}
\end{minipage}
\begin{minipage}[c]{0.13\textwidth}
\epsfig{file=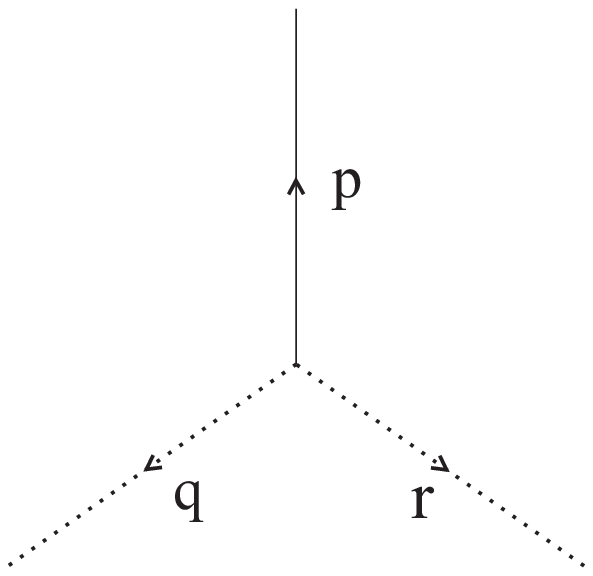,height=2cm}
\end{minipage}%
\begin{minipage}[c]{0.175\textwidth}
\small
\begin{equation*}
    \leftrightarrow  \tilde{\Gamma}^{t(0)}_{(0,1,0,2)}[p;q,r]
\end{equation*}
\end{minipage}
\begin{minipage}[c]{0.13\textwidth}
\epsfig{file=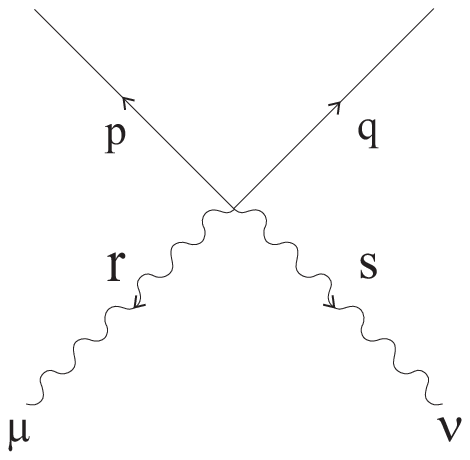,height=2cm}
\end{minipage}%
\begin{minipage}[c]{0.195\textwidth}
\small
\begin{equation*}
    \leftrightarrow  \tilde{\Gamma}^{t(0)\m\n}_{(2,2,0,0)}[r,s;p,q]
\end{equation*}
\end{minipage}
\begin{minipage}[c]{0.13\textwidth}
\epsfig{file=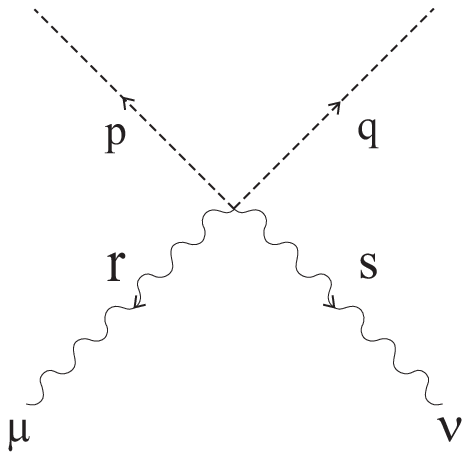,height=2cm}
\end{minipage}%
\begin{minipage}[c]{0.195\textwidth}
\small
\begin{equation*}
    \leftrightarrow  \tilde{\Gamma}^{t(0)\m\n}_{(2,0,2,0)}[r,s;p,q]
\end{equation*}
\end{minipage}
\begin{minipage}[c]{0.13\textwidth}
\epsfig{file=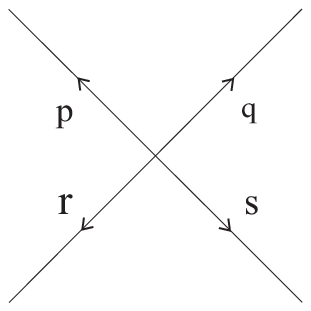,height=2cm}
\end{minipage}%
\begin{minipage}[c]{0.195\textwidth}
\small
\begin{equation*}
    \leftrightarrow  \tilde{\Gamma}^{t(0)}_{(0,4,0,0)}[p,q,r,s]
\end{equation*}
\end{minipage}
\begin{minipage}[c]{0.13\textwidth}
\epsfig{file=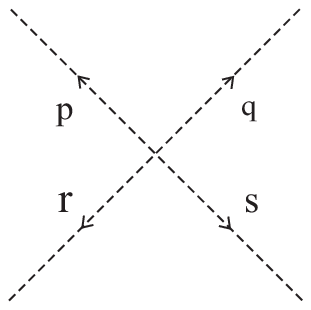,height=2cm}
\end{minipage}%
\begin{minipage}[c]{0.195\textwidth}
\small
\begin{equation*}
    \leftrightarrow \tilde{\Gamma}^{t(0)}_{(0,0,4,0)}[p,q,r,s]
\end{equation*}
\end{minipage}
\begin{minipage}[c]{0.13\textwidth}
\epsfig{file=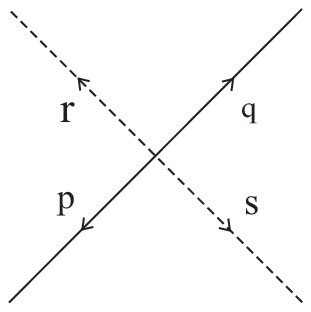,height=2cm}
\end{minipage}%
\begin{minipage}[c]{0.195\textwidth}
\small
\begin{equation*}
    \leftrightarrow  \tilde{\Gamma}^{t(0)}_{(0,2,2,0)}[p,q;r,s]
\end{equation*}
\end{minipage}
\begin{minipage}[c]{0.13\textwidth}
\epsfig{file=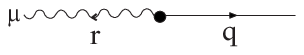,width=2cm}
\end{minipage}%
\begin{minipage}[c]{0.17\textwidth}
\small
\begin{equation*}
    \leftrightarrow \tilde{\Gamma}^{t(1)\m}_{(1,1,0,0)}[r;q]
\end{equation*}
\end{minipage}
\begin{minipage}[c]{0.13\textwidth}
\epsfig{file=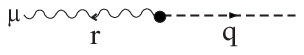,width=2cm}
\end{minipage}%
\begin{minipage}[c]{0.17\textwidth}
\small
\begin{equation*}
    \leftrightarrow \tilde{\Gamma}^{t(1)\m}_{(1,0,1,0)}[r;q]
\end{equation*}
\end{minipage}
\begin{minipage}[c]{0.13\textwidth}
\epsfig{file=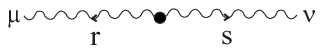,width=2cm}
\end{minipage}%
\begin{minipage}[c]{0.17\textwidth}
\small
\begin{equation*}
    \leftrightarrow \tilde{\Gamma}^{t(1)\m\n}_{(2,0,0,0)}[r,s]
\end{equation*}
\end{minipage}
\begin{minipage}[c]{0.13\textwidth}
\epsfig{file=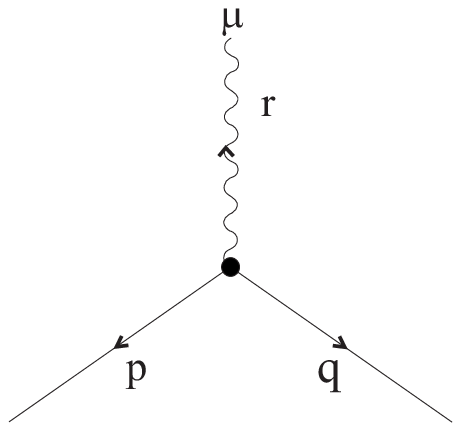,width=2cm}
\end{minipage}%
\begin{minipage}[c]{0.175\textwidth}
\small
\begin{equation*}
    \leftrightarrow \tilde{\Gamma}^{t(1)\m}_{(1,2,0,0)}[r;p,q]
\end{equation*}
\end{minipage}
\begin{minipage}[c]{0.13\textwidth}
\epsfig{file=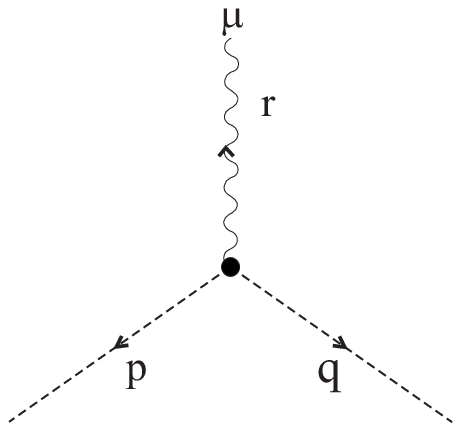,width=2cm}
\end{minipage}%
\begin{minipage}[c]{0.175\textwidth}
\small
\begin{equation*}
    \leftrightarrow  \tilde{\Gamma}^{t(1)\m}_{(1,0,2,0)}[r;p,q]
\end{equation*}
\end{minipage}
\begin{minipage}[c]{0.13\textwidth}
\epsfig{file=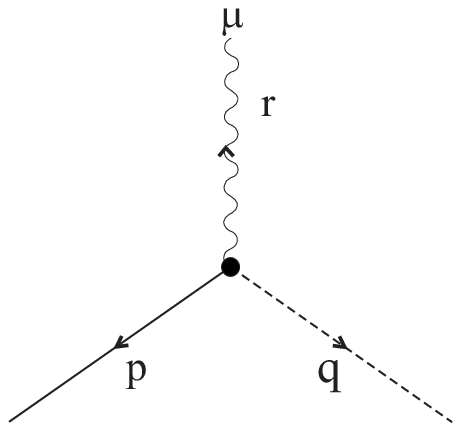,width=1.8cm}
\end{minipage}%
\begin{minipage}[c]{0.175\textwidth}
\small
\begin{equation*}
    \leftrightarrow \tilde{\Gamma}^{t(1)\m}_{(1,1,1,0)}[r;p;q]
\end{equation*}
\end{minipage}
\begin{minipage}[c]{0.13\textwidth}
\epsfig{file=verticeaaat.eps,width=1.8cm}
\end{minipage}%
\begin{minipage}[c]{0.175\textwidth}
\small
\begin{equation*}
    \leftrightarrow \tilde{\Gamma}^{t(1)\m\n\e}_{(3,0,0,0)}[p,q,r]
\end{equation*}
\end{minipage}
\begin{minipage}[c]{0.13\textwidth}
\epsfig{file=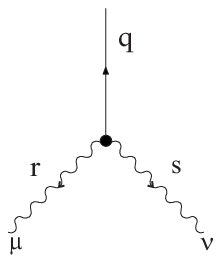,width=1.8cm}
\end{minipage}%
\begin{minipage}[c]{0.172\textwidth}
\small
\begin{equation*}
    \leftrightarrow  \tilde{\Gamma}^{t(1)\m\n}_{(2,1,0,0)}[r,s;q]
\end{equation*}
\end{minipage}
\begin{minipage}[c]{0.13\textwidth}
\epsfig{file=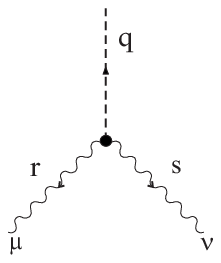,width=1.8cm}
\end{minipage}%
\begin{minipage}[c]{0.172\textwidth}
\small
\begin{equation*}
    \leftrightarrow \tilde{\Gamma}^{t(1)\m\n}_{(2,0,1,0)}[r,s;q]
\end{equation*}
\end{minipage}
\begin{minipage}[c]{0.13\textwidth}
\epsfig{file=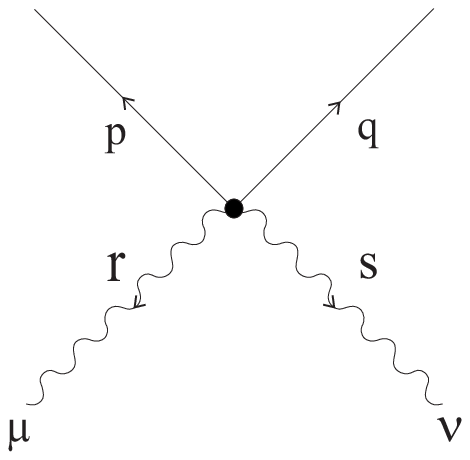,width=1.8cm}
\end{minipage}%
\begin{minipage}[c]{0.194\textwidth}
\small
\begin{equation*}
    \leftrightarrow \tilde{\Gamma}^{t(1)\m\n}_{(2,2,0,0)}[r,s;p,q]
\end{equation*}
\end{minipage}
\begin{minipage}[c]{0.13\textwidth}
\epsfig{file=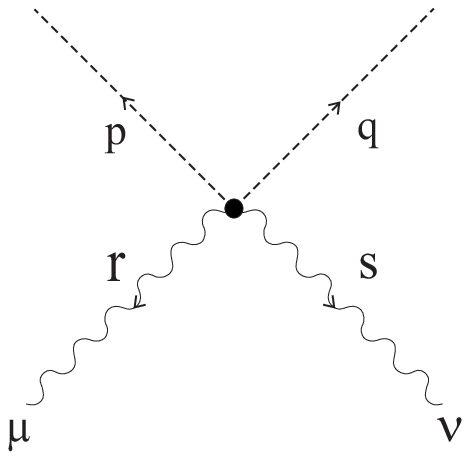,width=1.8cm}
\end{minipage}%
\begin{minipage}[c]{0.194\textwidth}
\small
\begin{equation*}
    \leftrightarrow  \tilde{\Gamma}^{t(1)\m\n}_{(2,0,2,0)}[r,s;p,q]
\end{equation*}
\end{minipage}
\begin{minipage}[c]{0.13\textwidth}
\epsfig{file=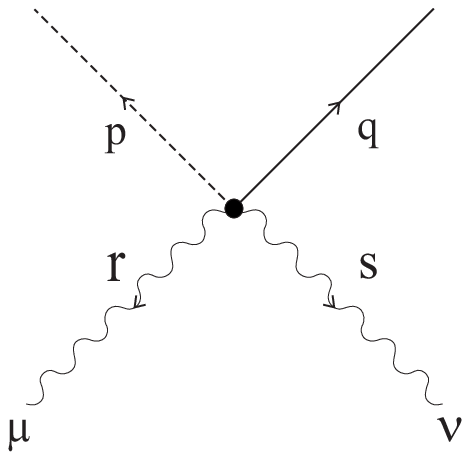,width=1.8cm}
\end{minipage}%
\begin{minipage}[c]{0.194\textwidth}
\small
\begin{equation*}
    \leftrightarrow \tilde{\Gamma}^{t(1)\m\n}_{(2,1,1,0)}[r,s;q;p]
\end{equation*}
\end{minipage}
\begin{minipage}[c]{0.13\textwidth}
\epsfig{file=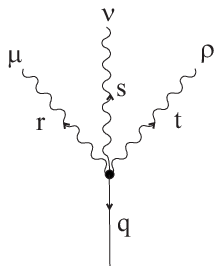,width=1.8cm}
\end{minipage}%
\begin{minipage}[c]{0.194\textwidth}
\small
\begin{equation*}
    \leftrightarrow \tilde{\Gamma}^{t(1)\m\n\r}_{(3,1,0,0)}[r,s,t;q]
\end{equation*}
\end{minipage}
\begin{minipage}[c]{0.13\textwidth}
\epsfig{file=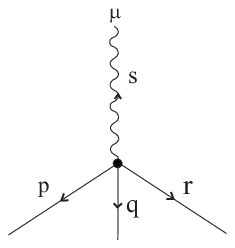,width=1.8cm}
\end{minipage}%
\begin{minipage}[c]{0.195\textwidth}
\small
\begin{equation*}
    \leftrightarrow \tilde{\Gamma}^{t(1)\m}_{(1,3,0,0)}[s;p,q,r]
\end{equation*}
\end{minipage}
\begin{minipage}[c]{0.13\textwidth}
\epsfig{file=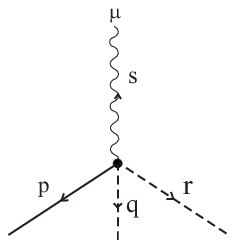,width=1.8cm}
\end{minipage}%
\begin{minipage}[c]{0.22\textwidth}
\small
\begin{equation*}
    \leftrightarrow \tilde{\Gamma}^{t(1)\m}_{(1,1,2,0)}[s;p;q,r]
\end{equation*}
\end{minipage}
\begin{minipage}[c]{0.13\textwidth}
\epsfig{file=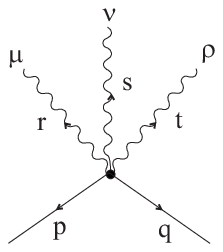,width=1.8cm}
\end{minipage}%
\begin{minipage}[c]{0.22\textwidth}
\small
\begin{equation*}
    \leftrightarrow \tilde{\Gamma}^{t(1)\m\n\r}_{(3,2,0,0)}[r,s,t;p,q]
\end{equation*}
\end{minipage}
\end{minipage}
\begin{minipage}{0.96\textwidth}
\centering
\begin{minipage}[c]{0.13\textwidth}
\epsfig{file=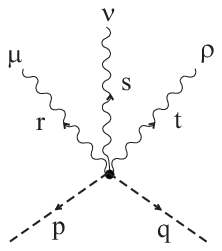,width=2cm}
\end{minipage}%
\begin{minipage}[c]{0.23\textwidth}
\small
\begin{equation*}
    \leftrightarrow \tilde{\Gamma}^{t(1)\m\n\r}_{(3,0,2,0)}[r,s,t;p,q]
\end{equation*}
\end{minipage}
\begin{minipage}[c]{0.13\textwidth}
\epsfig{file=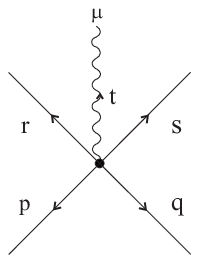,width=2cm}
\end{minipage}%
\begin{minipage}[c]{0.23\textwidth}
\small
\begin{equation*}
    \leftrightarrow \tilde{\Gamma}^{t(1)\m}_{(1,4,0,0)}[t;p,q,r,s]
\end{equation*}
\end{minipage}
\begin{minipage}[c]{0.13\textwidth}
\epsfig{file=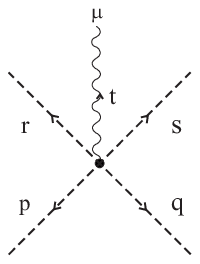,width=2cm}
\end{minipage}%
\begin{minipage}[c]{0.23\textwidth}
\small
\begin{equation*}
    \leftrightarrow  \tilde{\Gamma}^{t(1)\m}_{(1,0,4,0)}[t;p,q,r,s]
\end{equation*}
\end{minipage}
\begin{minipage}[c]{0.13\textwidth}
\epsfig{file=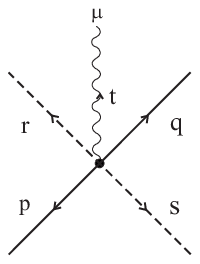,width=2cm}
\end{minipage}%
\begin{minipage}[c]{0.23\textwidth}
\small
\begin{equation*}
    \leftrightarrow  \tilde{\Gamma}^{t(1)\m}_{(1,2,2,0)}[t;p,q;r,s]
\end{equation*}
\end{minipage}
\\[13pt]
 \renewcommand{\figurename}{Fig.}
 \renewcommand{\captionlabeldelim}{.}
\figcaption{Feynman rules for the phase with spontaneously broken symmetry}
\end{minipage}
\\[\intextsep]
\begin{minipage}{0.95\textwidth}
  \centering
\epsfig{file=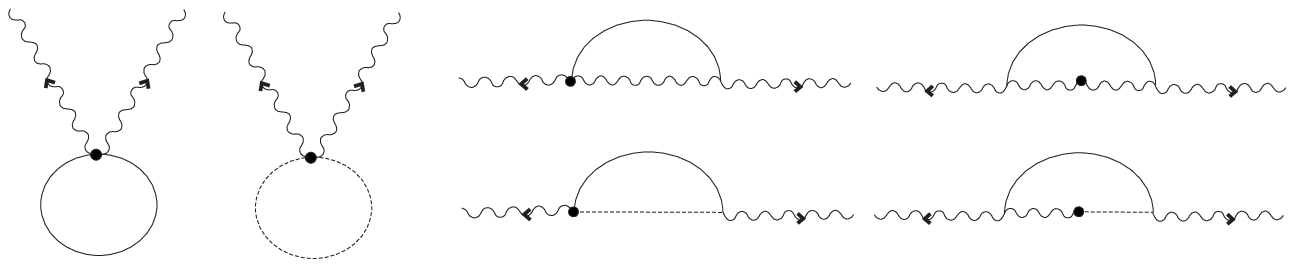,height=3cm}
\renewcommand{\figurename}{Fig.}
 \renewcommand{\captionlabeldelim}{.}
\\[13pt]
\figcaption{  Topologically inequivalent diagrams contributing to
the $M-$dependent part of the gauge field two-point function.}
 \end{minipage}
\\[\intextsep]
\begin{minipage}{0.95\textwidth}
  \centering
\epsfig{file=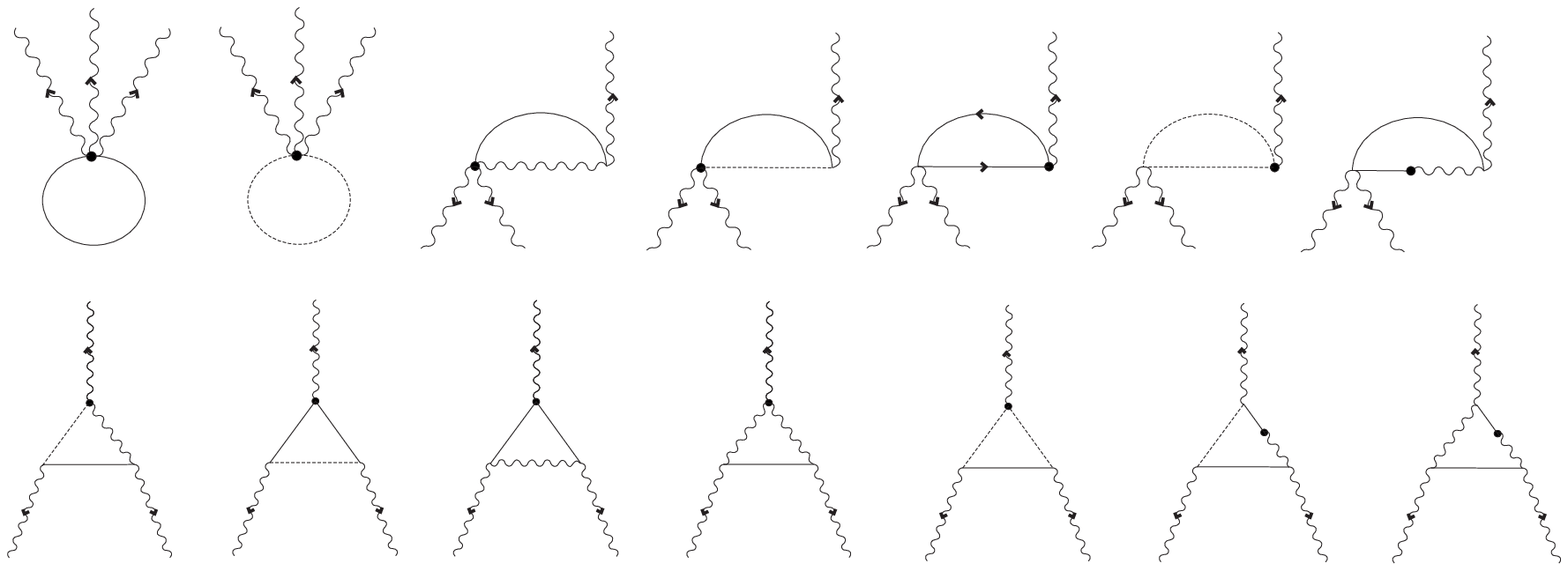,height=5.7cm}
\renewcommand{\figurename}{Fig.}
 \renewcommand{\captionlabeldelim}{.}
\figcaption{  Topologically inequivalent diagrams contributing to
the $M-$dependent part of the gauge field three-point function.}
 \end{minipage}

\end{document}